\documentclass[twocolumn]{revtex4-1}

\usepackage{amsmath, amssymb, array}
\usepackage[english]{babel}
\usepackage{balance}
\usepackage{bm}
\usepackage{booktabs}
\usepackage{color}
\usepackage{dcolumn}
\usepackage{epsfig}
\usepackage{float}
\usepackage{gensymb}
\usepackage{graphicx,rotating}
\usepackage[utf8]{inputenc}
\usepackage{mhchem}
\usepackage{multirow}
\usepackage{physics}
\usepackage[justification=centering]{subcaption}

\newcolumntype{C}[1]{>{\centering\arraybackslash}m{#1}} 

\begin{document}

\title{Optical properties of single-crystal As$_2$Se$_3$ from first principles}

\author{Juan J. Mel\'endez}
\affiliation{Department of Physics, University of Extremadura. Avda. de Elvas, s/n, 06006, Badajoz, Spain}
\affiliation{Institute for Advanced Scientific Computing of Extremadura, Badajoz, Spain}

\date{\today}

\begin{abstract}
	Crystalline As$_2$Se$_3$ is a promising layered chalcogenide for mid‑infrared photonics and optoelectronic applications, yet its fundamental electronic and optical properties remain debated. This paper combines density functional theory with G$_0$W$_0$ quasiparticle corrections and Bethe–Salpeter equation (BSE) calculations to deliver a robust description of its band structure and optical response. The G$_0$W$_0$ results reveal an indirect band gap of 2.31 eV with nearly degenerate direct transitions at $\Gamma$, while the BSE spectra exhibit strong polarization‑dependent anisotropy and pronounced excitonic effects. Several bound excitons are identified near the absorption onset, clarifying the role of electron–hole interactions and resolving previous controversies regarding the nature of the optical gap and the contribution of lone‑pair states. This work provides a unified and quantitatively accurate picture of As$_2$Se$_3$, reinforcing its potential for next‑generation infrared and anisotropic photonic devices.
\end{abstract}

\maketitle

\section{Introduction}
Metal chalcogenides have garnered sustained interest in materials science due to their extraordinary array of physical properties, ranging from unique optical and electronic characteristics to topological phases \cite{Furdyna-20}. These materials have found extensive utility in modern technologies, including phase-change memories, solar cells, and mid-infrared photonic devices \cite{Eggleton-11, Chung-14}. In particular, arsenic triselenide (As$_2$Se$_3$) has been the subject of fundamental study owing to its quasi-two-dimensional nature and high potential for optoelectronic applications. It crystallizes in the monoclinic system with space group $P2_1/n$ \cite{Renninger-73, Stergiou-85}, exhibiting a structural isomorphism with the mineral orpiment (As$_2$Se$_3$) \cite{Zallen-71}. The lattice is composed of puckered layers stacked along the unique monoclinic axis, formed by spiral chains of pyramidal AsSe$_3$ units linked by covalent bonds. Early investigations established that the van der Waals interaction between these layers, despite being weak, plays a crucial role in valence band dispersion and the macroscopic response of the material \cite{Zallen-71, Althaus-78}.

Although the vast majority of optical studies on this material have focused on its vitreous state due to its technological relevance as a glass former \cite{Kolomiets-64, Zallen-71, Eggleton-11, Fayek-07, Struzhkin-08}, experimental and theoretical works exist for the crystalline phase that reveal a complex electronic landscape. For the purpose of this paper, it may be highlighted that the absorption spectrum of crystalline As$_2$Se$_3$ is characterized by a fundamental absorption edge located at approximately 2.0 eV at room temperature \cite{Shaw-70, Althaus-78}. A distinctive feature of the optical response of this semiconductor is its pronounced anisotropy; the absorption coefficient and refractive index show significant variation depending on the light polarization relative to the crystal axes \cite{Shaw-70, Zallen-76, Althaus-78}. This anisotropy seems to be a direct consequence of the low-symmetry layered structure and the specific orbital contributions near the Fermi level, which are derived primarily from the hybridization of As-4p and Se-4p states \cite{Tarnow-86b}.

Despite extensive literature---from the pioneering work on electronic conduction by Kolomiets \cite{Kolomiets-64} to detailed optical absorption studies by Evans and Young \cite{Evans-67} and Zallen et al.\ \cite{Zallen-71}---fundamental discrepancies persist between theoretical models and experimental data. A central controversy concerns the nature of the valence electronic states. While the intuitive chemical model proposed by Kastner \cite{Kastner-72} and X-ray photoemission spectroscopy (XPS) measurements by Bishop and Shevchik \cite{Bishop-75} suggest that the valence band maximum is dominated by well-defined non-bonding p-orbitals (``lone pairs'') of Se, the tight-binding analysis by Althaus et al.\ \cite{Althaus-78} contradicts this view. Their calculations indicate that, in the crystalline phase, hybridization is sufficiently strong to eliminate the energetic separation of lone pairs by mixing bonding and non-bonding states, a factor that complicates the interpretation of reflectivity spectra.

Furthermore, there is a lack of consensus regarding the exact nature of the fundamental absorption edge. Most studies conclude that As$_2$Se$_3$ is an indirect gap semiconductor \cite{Althaus-78, Blabla-76}, although a direct transition at $\Gamma$ is identified as well. By contrast, Zallen and co-workers discuss strong evidence pointing out that the gap in this semiconductor is actually direct \cite{Zallen-71, Zallen-76}. This ambiguity is exacerbated by the findings of Tarnow et al.\ \cite{Tarnow-86b}, who demonstrated via total energy calculations that interlayer interaction generates non-zero off-diagonal elements in the dielectric tensor, which implies that the optical axes rotate as a function of frequency. This is a phenomenon that simplified models cannot capture, which may be the reason underlying the fact that the nature of the absorption edge in this semiconductor has not been clearly characterized so far. 

In recent decades, theoretical efforts have shifted towards \textit{ab initio} computational approaches based on Density Functional Theory (DFT) to provide a more rigorous description of the material. Recently, Sharma et al.\ \cite{Sharma-12} employed the Full-Potential Linearized Augmented Plane Wave (FP-LAPW) method to analyze the partial density of states and frequency-dependent optical constants, confirming the dominant contribution of Se p-orbitals to the valence and conduction bands. Besides, studies by Tse et al.\ \cite{Tse-22} have extended this characterization to evaluate the structural, mechanical, and elastic stability of the As$_2$Se$_3$ lattice. However, despite these advances, the methods used to date present systematic limitations. It is well known that standard DFT functionals severely underestimate the fundamental band gap, often requiring empirical ``scissors'' corrections to align with experiment. More critically, these single-particle approaches fail to account for many-body electron-hole interactions (excitons), which are known to significantly reshape the optical spectra of chalcogenide crystals near the absorption edge.

In the light of the previous, this work aims a double goal, namely \textit i) characterize the band structure of As$_2$Se$_3$ from a sound theoretical basis which overcome the aforementioned controversies, and \textit {ii}) characterize the optical properties of the system and correlate them with the band structure, with particular focus on the possible excitonic activity. To achieve these objectives, I implement here an advanced theoretical approach combining DFT with Many-Body Perturbation Theory (MBPT). This strategy employs the non-consistent G$_0$W$_0$ formalism to correct quasiparticle energies and the Bethe-Salpeter Equation (BSE) to describe optical excitations. This combination allows us to unequivocally determine the direct or indirect nature of the gap, clarify the role of lone pairs in the optical spectrum, and provide a precise quantitative description of the optical properties of As$_2$Se$_3$ that unifies divergent experimental observations. 

\section{Methodology}
DFT calculations were carried out by the Quantum Espresso package \cite{Giannozzi-09} using the projector augmented wave (PAW) method \cite{Blochl-94}. The generalized-gradient approximation (GGA) with the Perdew-Burke-Ernzerhof parametrization was adopted to describe the exchange-correlation functional \cite{Perdew-96}. After convergence tests to within 0.001 eV, the kinetic energy cutoff for the wave functions was set to 40 Ry, and a Monkhorst-Pack $k$-mesh of $6\times6\times2$ \cite{Monkhorst-76} was used to sample the first Brillouin zone (BZ) at the self-consistent stage. 

The starting monoclinic cell (space group $P2_1/c$) with the experimental lattice parameters $a =$~4.30~\AA, $b=$~9.94~\AA, $c=$~12.84~\AA, $\beta=$ 109.1º \cite{Renninger-73} was relaxed following the BFGS method\footnote{Note that in a previous work \cite{Gonzalez-18} a different setting was chosen for the lattice vectors of the monoclinic lattice.
}; the details of the geometry optimization of the unit cell were reported elsewhere \cite{Gonzalez-18}. For both the geometry optimization and the subsequent self-consistent calculations at the relaxed structure, van der Waals corrections modeled with the DFT-D method by Grimme et al. were explicitly used \cite{Grimme-10} since, as mentioned in the introduction, interlayer interactions play a role in the electronic structure of the solid. The band structure was calculated along the $Z \left(0, \frac 12, 0\right) \rightarrow C \left(\frac 12, \frac 12, 0\right) \rightarrow \Gamma \rightarrow B \left(0, 0, \frac 12\right) \rightarrow D\left(0, \frac 12, \frac 12\right) \rightarrow \Gamma \rightarrow Y\left(-\frac 12, 0, 0\right) \rightarrow A \left(-\frac 12, 0, \frac 12\right) \rightarrow E \left(-\frac 12, \frac 12, \frac 12\right) \rightarrow \Gamma$ path after band interpolation using the Wannier90 code \cite{Marzari-97}. 

MBPT corrections were made from the DFT solutions following the non-consistent $G_0W_0$ scheme. After the convergence analysis to within 0.01 eV, a cutoff of 9 Ry was set for the exchange self-energy and a 6$\times$6$\times$3 Monkhorst-Pack grid was used to sample the first Brillouin zone. The plasmon-pole approximation \cite{Godby-89} was used to compute the electronic screening by summing 400 bands with a cutoff of 3 Ry, and 300 bands were used for the $G_0W_0$ summation. The optical spectrum was computed from the MBPT band structure by solving the BSE with the static part of the previously computed PPA dielectric matrix and using a 1 Ry cutoff for the screened and exchange interactions; 19 bands (namely the 6 topmost valence bands and the 13 bottommost conduction bands) were considered in the calculation of the absorption spectra. These were computed for three orientations (namely parallel to the $OX$, $OY$ and $OZ$ axes) of the electric field. All MBPT and BSE calculations were carried out using the Yambo code \cite{Marini-09}. The complete details about the convergence of the several G$_0$W$_0$ and BSE parameters may be found in the Supplement.

\section{Results}
\subsection{Geometry optimization}
\begin{figure}[!h]
	\centering
	\includegraphics[width=\columnwidth]{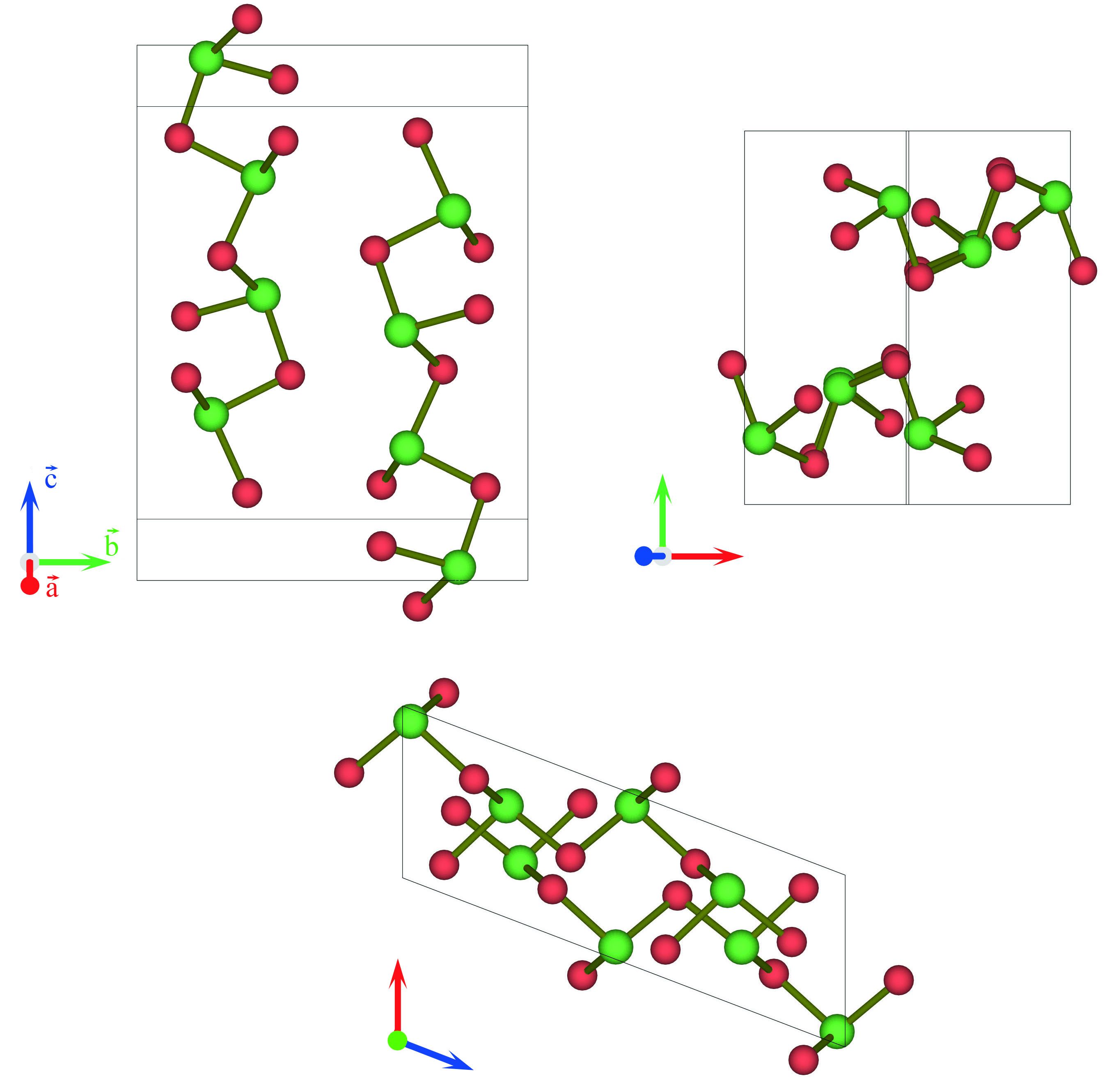}
	\caption{Views of the unit cell of As$_2$Se$_3$ along the crystal axes after the structural relaxation.}
	\label{Fig_1}
\end{figure}

The relaxed crystal structure of As$_2$Se$_3$ is shown in Fig.~\ref{Fig_1}. It consists of two ``layers'', each comprising two As$_2$Se$_3$ units, stacked along the $\vec b$ axis (as shown in Fig. \ref{Fig_1} top left) and shifted about $\frac 14$ along $\vec a$. This pseudo-layered structure is similar to that of other chalcogenides such as SnSe, and justifies the explicit use of dispersive corrections, as the interlayer interactions are expected to be low \cite{Zallen-76}. After relaxation, the calculated cell parameters were $a = 4.74$~\AA, $b = 10.78$~\AA, $c = 13.07$~\AA~and $\beta = 110.92$º, which show a deviation of around 10\% with respect to the experimental values except for the $c$ axis, which has a deviation of around 2\%. These deviations are much higher than those commonly found in DFT, regardless of the exchange-correlation functional used, which is somehow surprising. Unfortunately, I do not have a concluding explanation for these abnormal discrepancies. One possible explanation could lie on how the software compute forces and stress tensor and, more especifically, on the accuracy of these calculations, especially in non-orthogonal symmetries. In the end, forces and stress tensor rule the BFGS procedure, and could result in abnormally high deviations in monoclinic systems. 

\subsection{Band structure and density of states}
\begin{figure}[!h]
	\centering
	\includegraphics[width=\columnwidth]{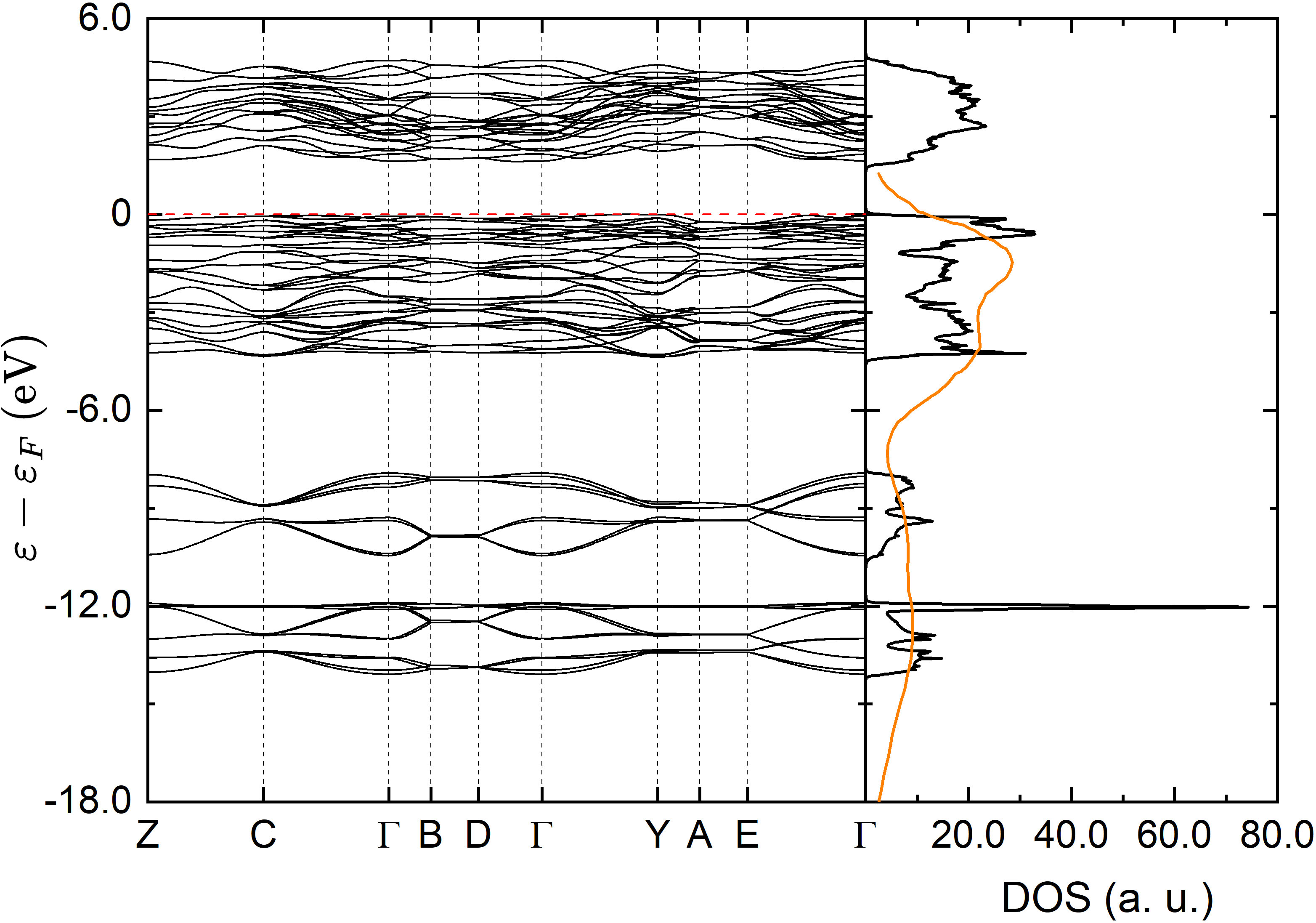}
	\caption{DFT band structure along high-symmetry lines of the Brillouin zone (left) and DoS (right) of As$_2$Se$_3$. The orange solid line corresponds to X-rays photoemission data by Bishop and Shevchik \cite{Bishop-75}.}
	\label{Fig_2}
\end{figure}

Fig. \ref{Fig_2} shows the DFT band structure (left) as well as the density of states (DoS, right) of As$_2$Se$_3$; for convenience, the energy origin has been set at the top of the valence band. 

The band structure exhibits three distinctive valence-band groups. The bottommost one consists of twelve bands and lies in the -14.0 to -12.0~eV energy range. There is a second one comprising eight bands with energies ranging between -10.0 and -8.0~eV. The topmost valence-band group extends from -4.4~eV and contains 36 bands. The calculated DoS shows a reasonably good agreement with data from X-ray photoemission reported by Bishop and Shevchik \cite{Bishop-75}, shown as a solid orange line, in both width of the signal and positions of the maxima. 

\begin{figure}[!h]
	\centering
	\includegraphics[width=\columnwidth]{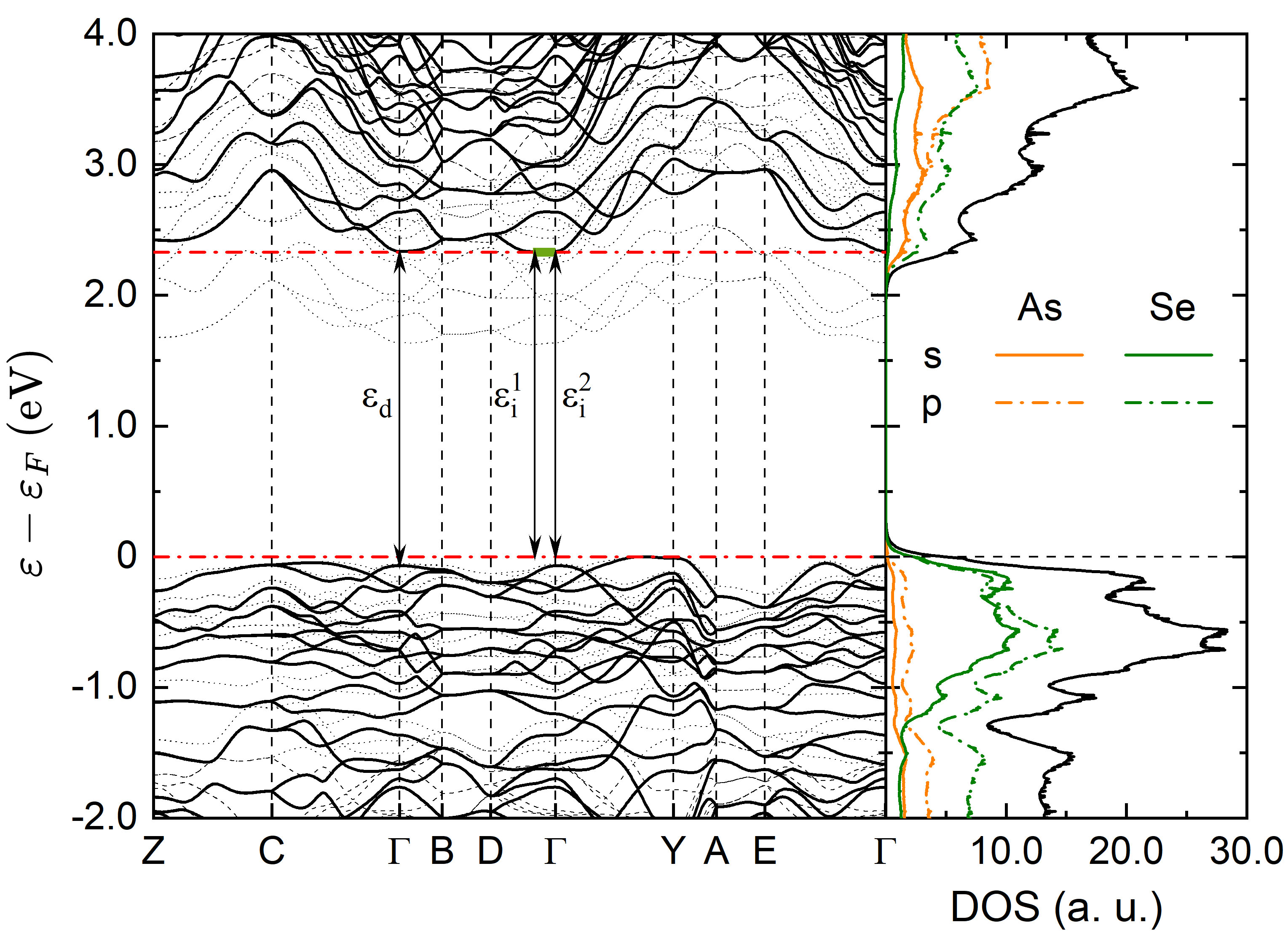}
	\caption{G$_0$W$_0$ band structure near the gap (left, solid line), and DoS (right) of As$_2$Se$_3$. The dashed lines correspond to the band structure calculated at the DFT level. The projections of the DoS onto s and p states of As and Se is also shown. }
	\label{Fig_3}
\end{figure} 

The DFT gap, $\varepsilon^i =$ 1.61~eV, is indirect, with the valence band maximum (VBM) along the $\Gamma - Y \left(\frac 12, 0, 0\right)$ path, close to $Y$, and the conduction band minimum (CBm) located at $\Gamma$. This result is consistent with a local orbital calculation by Bullet \cite{Blabla-76}, perhaps fortuitously, but it is below the 1.85~eV gap reported by Althaus et al. \cite{Althaus-78} and well above DFT data by Tarnow et al. \cite{Tarnow-86a}, who report a flat indirect gap of around 1.3 ~eV (to within 0.1 eV). Also the locations of the extremes differ with the later study, as Tarnow et al. find the VBM around the $\Gamma - B\left(0,0,\frac 12\right)$ line\footnote{I remark here that Tarnow et al. did use a different setting, namely that used in Ref.~ \cite{Gonzalez-18}.} (note that $B$ and $Y$ are nonequivalent for this cell) and the CBm at several locations within BZ, one of which being $\Gamma$. Interestingly, both the valence and the conduction bands are quite flat around their respective extremes, i.e., along the $\Gamma - Y$ and $\Gamma - D$ directions, respectively. Note that the conduction band exhibits a secondary minimum at $\Gamma$ around 0.01~eV larger than $\varepsilon^i$. This same finding was reported by previous works about the band structure of As$_2$Se$_3$ \cite{Althaus-78, Tarnow-86a, Blabla-76}.

In any case, the DFT results reported herein are much lower than the gap estimated for As$_2$Se$_3$ from optical measurements, which ranges between 2.0 and 2.2~eV \cite{Althaus-78, Shaw-70, Torane-03, Zallen-76}. To correct this discrepancy one must go beyond the bare DFT, which is well known to underestimate bandgaps in whichever of its flavors, to carry out many-body perturbative methods; in the case of the present work, I have performed G$_0$W$_0$ calculations within the PPA, as explained in the Methodology section. Fig.~\ref{Fig_3} left shows a detail of the G$_0$W$_0$ band structure near the gap. For comparison purposes, the DFT bands remain as dotted lines. This plot shows that the G$_0$W$_0$ calculations preserve the locations of the bands extremes but significantly widens the band gap. The indirect gap ($\varepsilon_i^1$ in Fig.~\ref{Fig_3}) is estimated in 2.31~eV, well above the DFT gap. The G$_0$W$_0$ calculations also preserve the overall band topology, like the non-dispersive behavior of the valence band along some directions; this fact is highlighted by the thick green band in Fig.~\ref{Fig_3}. In fact, there is a secondary VBM located exactly at $\Gamma$ ($\varepsilon_i^2$ in Fig.~\ref{Fig_3}) which differs from $\varepsilon_i^1$ by less than 0.01~eV, a difference which is within the margin of uncertainties of the DFT+G$_0$W$_0$ calculations performed here. For the subsequent discussion of optical properties, I remark here that the smallest direct gap, on the other hand, is $\varepsilon_d =$ 2.40~eV, and appears at $\Gamma$.

Fig.~\ref{Fig_3} right shows the density of states around the G$_0$W$_0$ gap, together with the projection of the DoS onto the atomic states of As and Se. The shape of the DoS is relatively sharp around the gap edges and especially at the top of the valence band. This fact results from the aforementioned flat landscape of the bands, and suggests strong optical transitions regardless of their nature, since both the initial and the final states are dense. Besides, while the edge of the conduction bands is highly hybridized, the top of the valence band arises mostly by the superposition of Se-s and Se-p states. I will come back to this issue in the discussion below.

\subsection{Optical properties}
\begin{figure}[H]
	\centering
	\begin{subfigure}[b]{\columnwidth}
		\centering
		\includegraphics[width=0.75\textwidth]{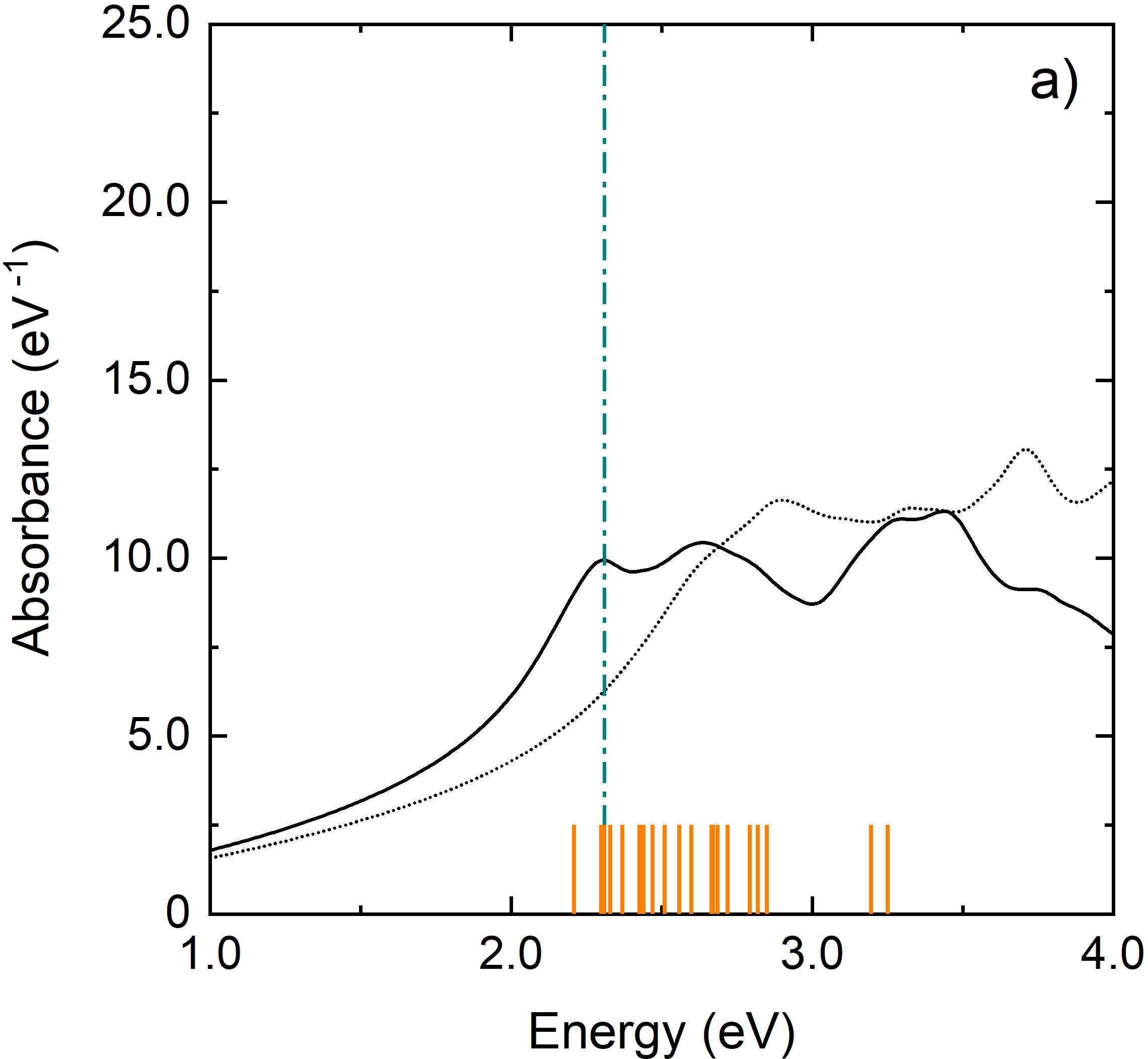}
		\caption{$E_x$}
		\label{fig:Exciton_Ex}
	\end{subfigure}
	
	\begin{subfigure}[b]{\columnwidth}
		\centering
		\includegraphics[width=0.75\textwidth]{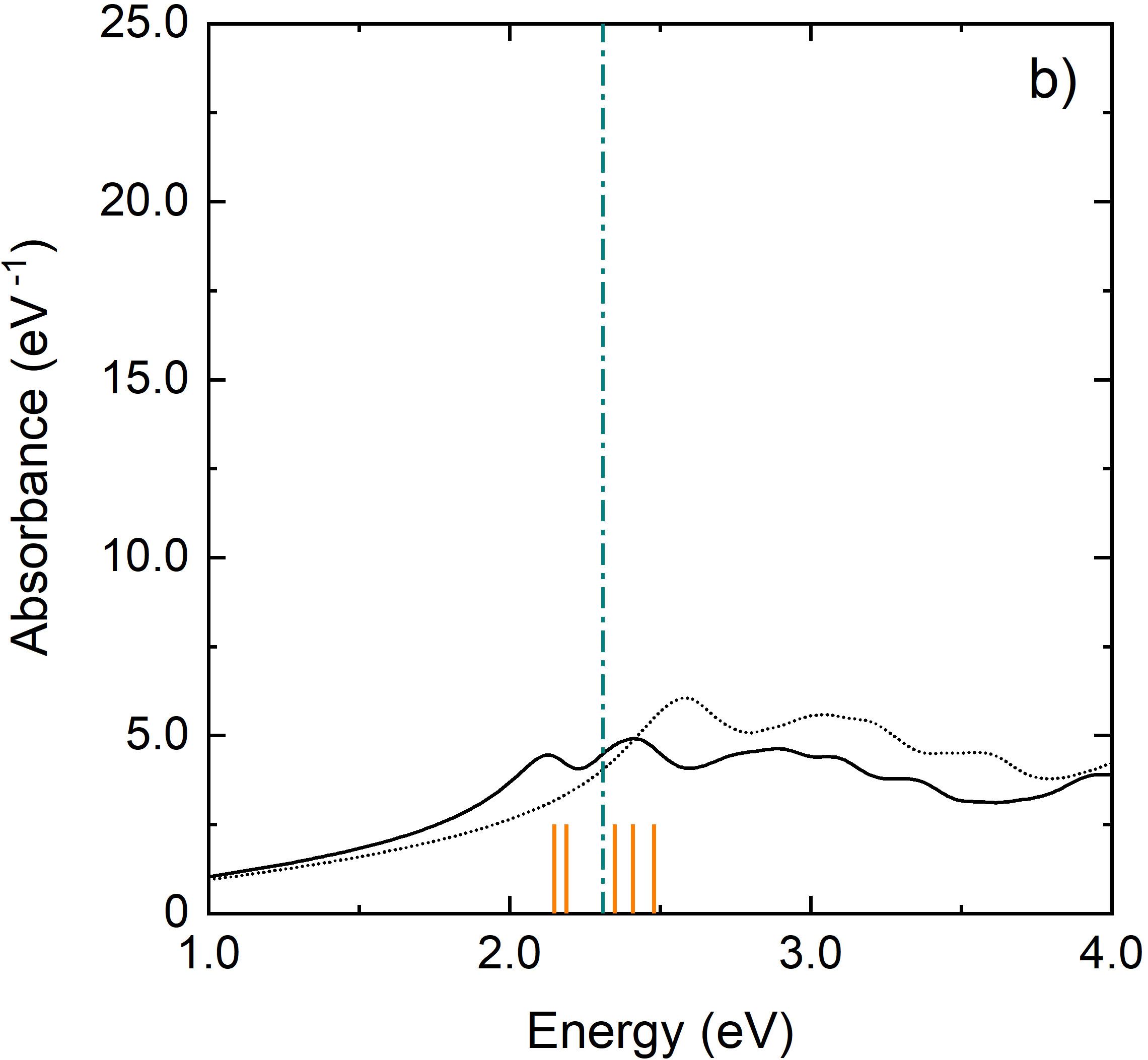}
		\caption{$E_y$}
		\label{fig:Exciton_Ey}
	\end{subfigure}   
	
	\begin{subfigure}[b]{\columnwidth}
		\centering
		\includegraphics[width=0.75\textwidth]{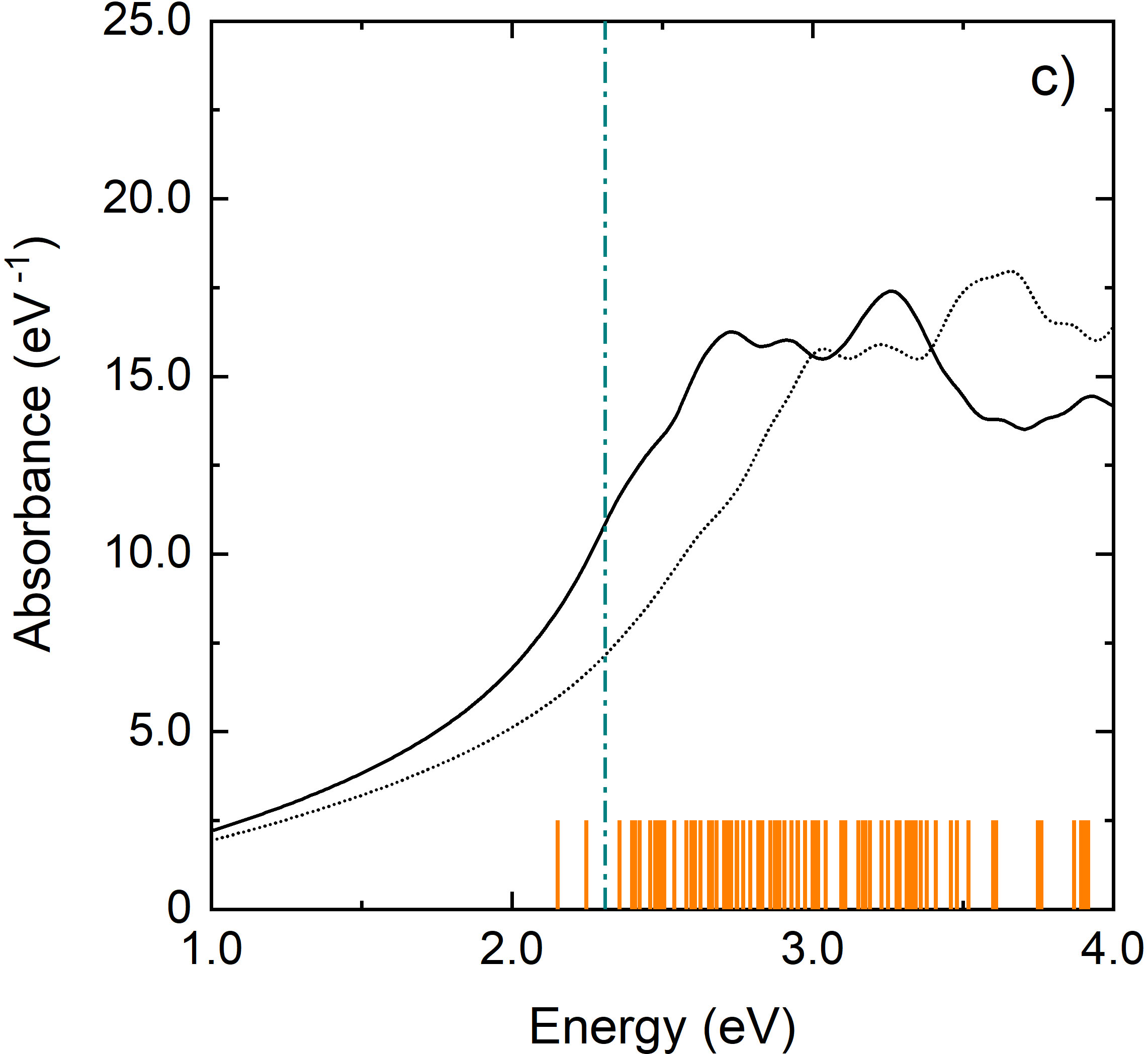}
		\caption{$E_z$}
		\label{fig:Exciton_Ez}
	\end{subfigure}
	\caption{Absorption spectra for As$_2$Se$_3$ computed by solving the BSE for three different orientations of the electric field (solid lines). Data computed within the IP approximation are also included for completeness (dotted lines). The vertical dashed line highlights the G$_0$W$_0$ gap, and the short solid lines the energetic positions of the excitons formed in each case (see the text for details). }
	\label{fig:Fig_4}
\end{figure}

As mentioned in the Methodology section, the G$_0$W$_0$ calculations were used as starting point to solve the BSE in search for possible excitonic behavior and aiming to clarify the nature of the optical transitions at the onset of the absorption spectrum. Figs.~\ref{fig:Fig_4} show the absorbance vs. energy curves for As$_2$Se$_3$ computed for the electric field aligned along the three Cartesian axes (solid lines); for completeness, the results obtained within the independent-particle (IP) approximations are also included for the same field polarizations as dotted lines. As I remarked before, not the whole set of bands entered into the calculations to solve the BSE, but only a few of them close to the bands extremes. In consequence, one should not extend the analysis of the absorption spectra much beyond a few eV over their onsets. 

The spectra in Figs.~\ref{fig:Fig_4} reveal first a marked anisotropy, as one could expect given not only the dissimilar lattice parameters of As$_2$Se$_3$, but also the characteristic symmetries of the space group. Thus, one notes first that absorption is enhanced for $\vec E \parallel OZ$, whereas it is minimum for $\vec E \parallel OY$ (for comparison purposes, the vertical axes in Figs.~\ref{fig:Fig_4} share the same scale). Also the spectrum onset and location of the absorption maxima near it depend on the orientation of the electric field. In particular, for $\vec E \parallel OX$ the first maximum appears roughly at the $G_0W_0$ gap, while for $\vec E \parallel OY$ and $\vec E \parallel OZ$ appear below (at $\varepsilon = 2.15$ eV) and above (at $\varepsilon = 2.65$~eV) it, respectively. This fact may be thought of as a possible sign of excitonic activity. 

Also, one notices that the BSE spectrum differs quite appreciably from that predicted by the IP approximation, especially for the field along $OX$ and $OZ$, i.e., those directions defining the $\beta \neq 90$º angle. More importantly, the BSE spectra are red shifted with respect to the IP spectrum in all cases, more importantly for the $OX$ and $OZ$ polarizations as well. Some peaks are evident in the BSE spectra which are absent (or obscured at least) in the IP spectra: note, for instance, the peak to the left of the vertical line for the $OY$ polarization, or the shoulder for the $OZ$ polarization close to the G$_0$W$_0$ gap. Both facts are also indicative of the possible excitonic activity in As$_2$Se$_3$.

The analysis of the solutions of the BSE confirms this idea showing that excitons form for all the field orientations. The positions of the bright (i.e., with non-zero oscillator strengths) exciton in As$_2$Se$_3$ are shown as vertical orange lines in Figs.~\ref{fig:Fig_4}; for simplicity, I have reported only those excitons whose intensity is greater than 0.2$I_0$, being $I_0$ the maximum exciton intensity. The number and location of these excitons are consistent with the aforementioned characteristics of the absorption spectra for each field orientation. Incidentally, it is remarkable that the number of excitons is high regardless of the polarization of the electric field. Although I will not extend the discussion further about this issue, some of these excitons actually coalesce, a well-known characteristics of indirect-gap semiconductors \cite{Zallen-76}.

Since typically the first (i.e., that appearing at the lowest energy) and most intense excitons in As$_2$Se$_3$ are those affecting more the optical properties, Table~\ref{table:excitons} includes some relevant data about these excitons, namely their position in the absorption spectrum, binding energy, intensity relative to the maximum one and bands involved, for the three orientations of the electric field. For completeness, Fig. S5 in the Supplement plots the exciton distribution in the reciprocal state (i. e., the projection of the excitonic states onto DFT states) for the excitons recorded in Table~\ref{table:excitons}. From this plot, one observes that the first exciton appears at $\Gamma$ regardless of the orientation of the electric field, and its relatively high binding energy is indicative of a high localization. The location of the most intense exciton, on the other hand, depends on the orientation of the electric field, appearing at the $B\left(0,0,\dfrac 12\right)$ point of the BZ for $\vec E \parallel OX$, at $\Gamma$ for $\vec E \parallel OY$ and at many locations within BZ for $\vec E \parallel OZ$; interestingly, the most intense exciton never appears at $Y$, where the VB takes on its maximum value. Note also that the relative intensity of the first exciton is less than 50\% that of the maximum, except for the $OY$ polarization. 

\begin{table*}[h!]
	\centering
	\renewcommand\arraystretch{1.5}
	\begin{tabular}{c c c c c c c }
		\toprule
		\hline
		\multicolumn{2}{c}{} & \textbf{$\boldsymbol{\varepsilon}$ (eV)} &\textbf{$\boldsymbol{\varepsilon_{bind.}}$ (eV)} & \textbf{$\boldsymbol{I/I_0}$} & \textbf{VB} & \textbf{CB} \\ 
		\hline
		\hline		
		\multirow{3}{*}{$E_x$} & First & 2.21 & 0.1 & 0.4 & 0 & 0 \\ \cline{4-7}
		& \multirow{2}{*}{Most intense} & \multirow{2}{*}{2.29} & \multirow{2}{*}{0.02} &\multirow{2}{*}{1} & {$-1$} & 0 \\ \cline{6-7} 
		& & & & & 0 & 0 \\ \hline
		
		\multirow{2}{*}{$E_y$} & First \& &\multirow{2}{*}{2.15} & \multirow{2}{*}{0.16} & \multirow{2}{*}{1}& \multirow{2}{*}{0} & \multirow{2}{*}{0} \\
		& most intense & & & & & \\ \hline 
		
		\multirow{3}{*}{$E_z$} & First & 2.15 & 0.16 & 0.16 & 0 & {0} \\ \cline{4-7}
		& \multirow{2}{*}{Most intense} &\multirow{2}{*}{2.94} & \multirow{2}{*}{Unbound} & \multirow{2}{*}{1}& {$-1$} & 0 \\ \cline{6-7} 
		& & & & & 0 & 0 \\ \hline
		\bottomrule
	\end{tabular}
	\caption{Characteristics of the first and most intense excitons in As$_2$Se$_3$ for three polarizations of the electric field. The VB and CB columns record the valence and conduction (respectively) bands where the corresponding excitons form; the index $-$1 indicates that the exciton forms in the next-to-last VB.}
	\label{table:excitons}
\end{table*} 

\section{Discussion}

\begin{figure*}[t]
	\centering
	\begin{subfigure}{\columnwidth}
		\centering
		\includegraphics[width=0.9\textwidth]{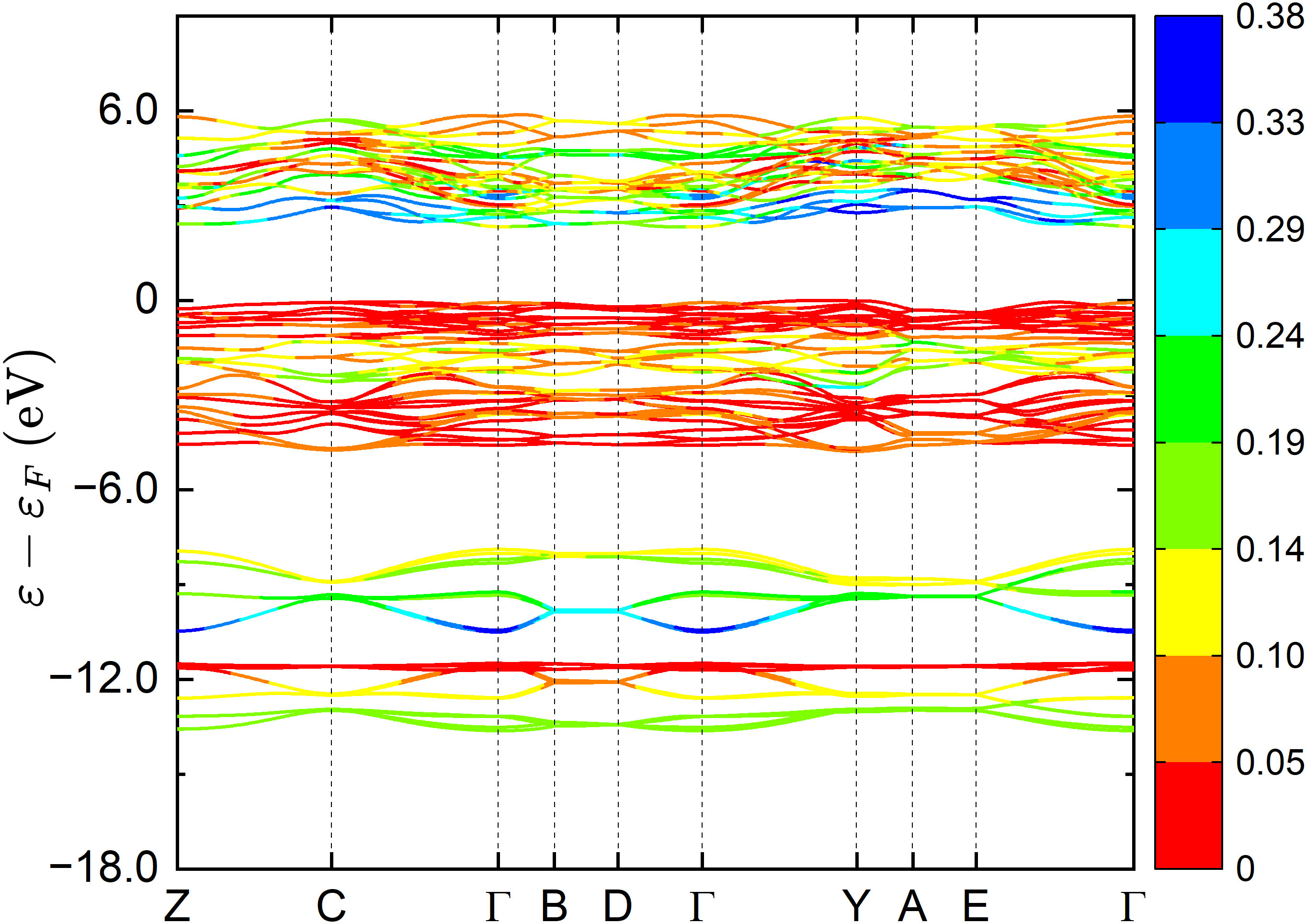}
		\caption{As-s states}
		\label{fig:Proj_As_s}
	\end{subfigure}
	\hfill
	\begin{subfigure}{\columnwidth}
		\centering
		\includegraphics[width=0.9\textwidth]{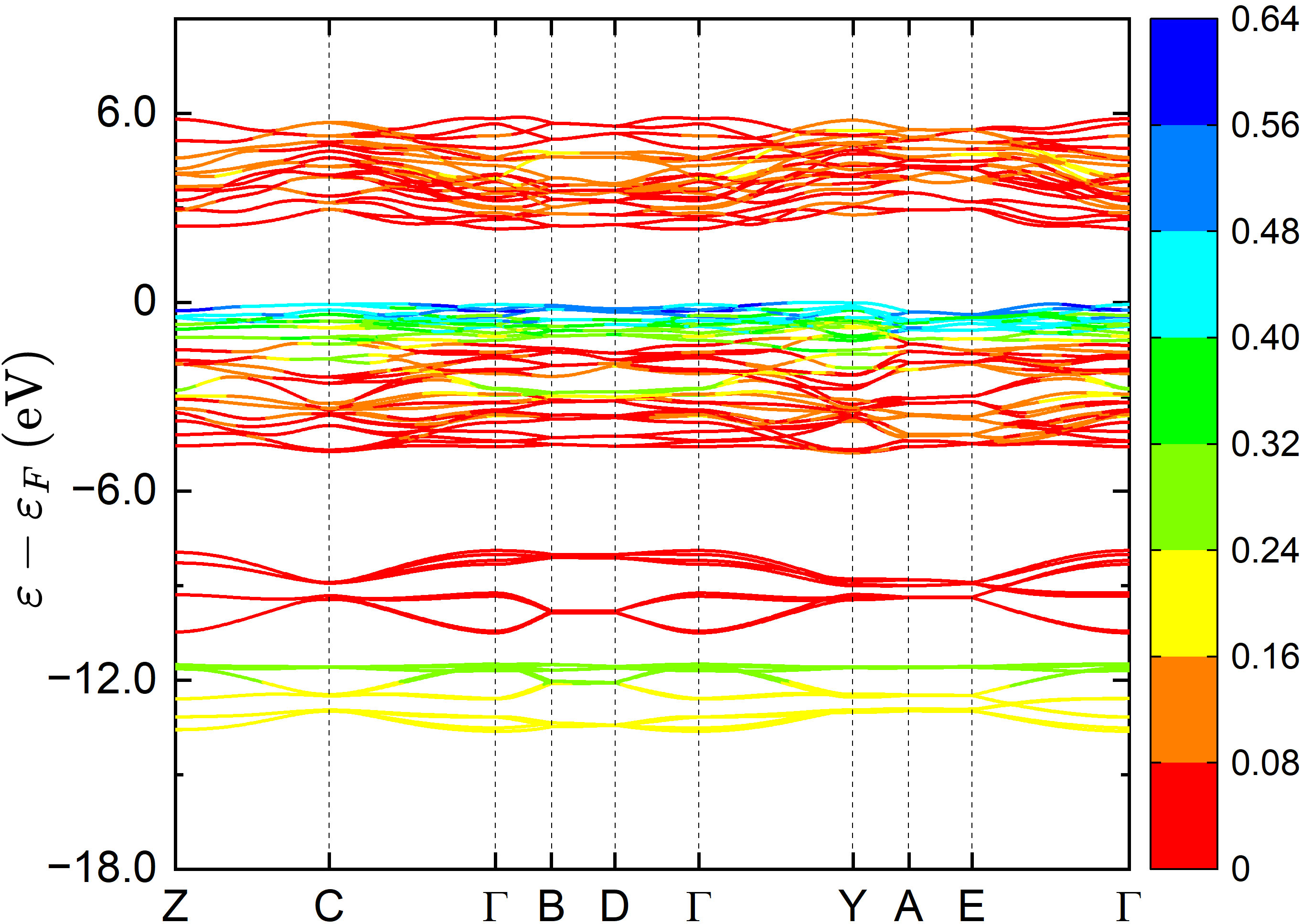}
		\caption{Se-s states}
		\label{fig:Proj_Se_s}
	\end{subfigure} 
	
	\qquad
	
	\begin{subfigure}{\columnwidth}
		\centering
		\includegraphics[width=0.9\textwidth]{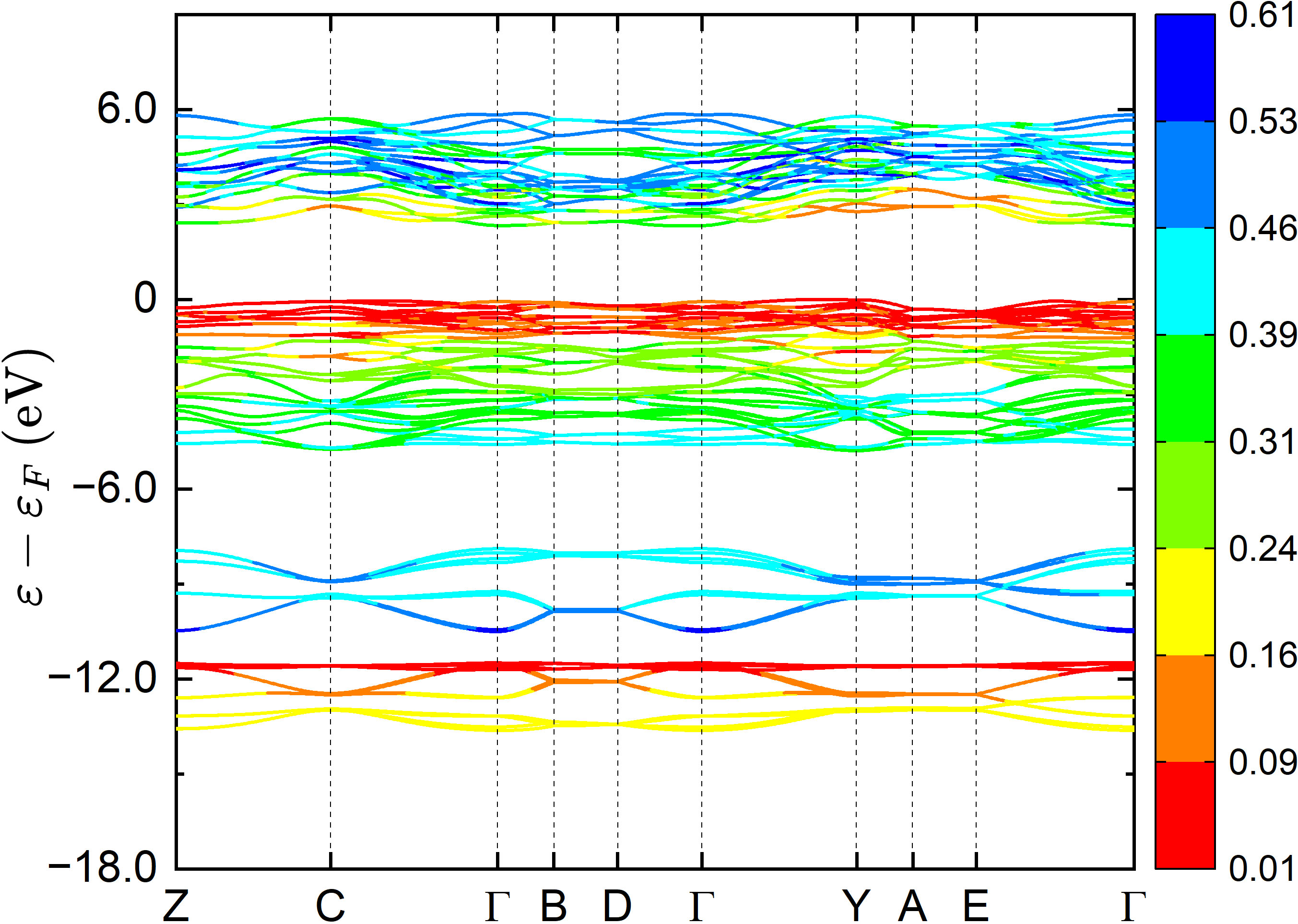}
		\caption{As-p states}
		\label{fig:Proj_As_p}
	\end{subfigure}   
	\hfill
	\begin{subfigure}{\columnwidth}
		\centering
		\includegraphics[width=0.9\textwidth]{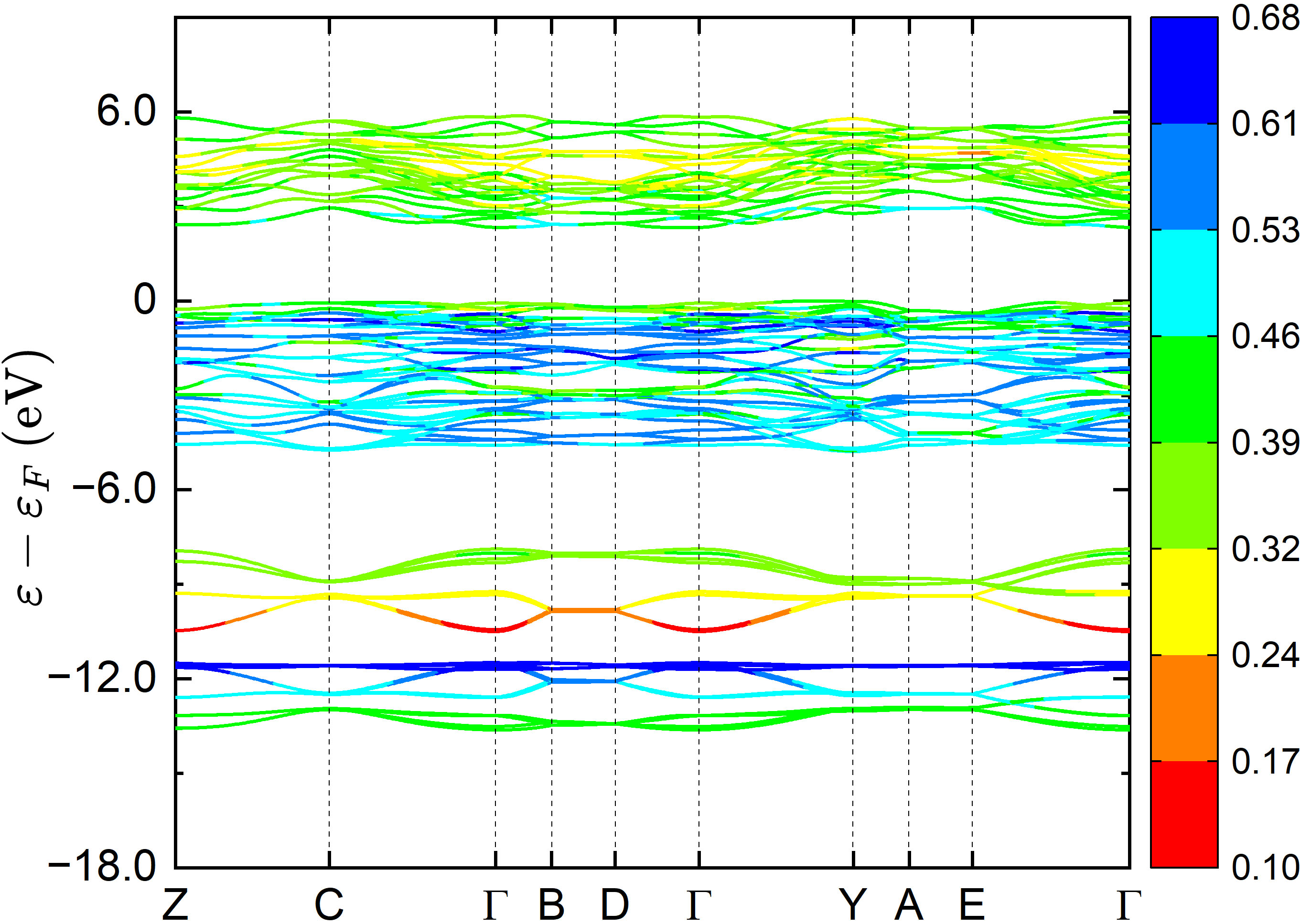}
		\caption{Se-p states}
		\label{fig:Proj_Se_p}
	\end{subfigure}
	\caption{Projections of the DFT band structure onto atomic states of As and Se.}
	\label{fig:projections}
\end{figure*}

The band structure of As$_2$Se$_3$ has received little attention in the past \cite{Blabla-76, Althaus-78, Tarnow-86a}. Althaus et al. \cite{Althaus-78} used a semi-empirical tight-binding model to calculate this band structure; even though the focus was put on the optical properties, their discussion of the band structure is quite complete. In particular, they discussed the arrangement of the atomic electrons within the bands. Specifically, the twelve bands comprising the bottommost group of bands were associated to the 24 Se-4$s$ electrons, and the next group to the 16 As-4$s$ ones. From DFT within the LDA approximation, Tarnow et al. came to the same conclusion \cite{Tarnow-86a}. The results of this paper contradict these pictures, however. This may be clarified by plotting the DFT band structure onto the s and p states of As and Se; these projections are shown in Fig.~\ref{fig:projections}. According to this figure, our results suggest instead that the bands within the two bottommost groups (energies below -8.0~eV) arise mostly from hybridization of both Se-4s and As-4s orbitals (see Fig.~\ref{fig:projections}). Some contribution from the As-4p ones exists at the sharp peak around -12.0~eV; in this sense, this level recalls an ``atomic-like'' (i.e., flat, no wave vector-dependent) one. The hybridization between As and Se $s$ states is even more balanced for the second group. 

Part of the discrepancies between our results and those reported earlier may be caused by the different methodologies used. In particular, Althaus et al. \cite{Althaus-78} did not use DFT, but a semiempirical tight-binding model, to study the band structure of As$_2$Se$_3$. The model used by the authors does not consider explicitly the electronic screening in the crystal field integrals; screening is indirectly accounted for by an adjustable parameter. Also, the authors use the atomic wavefunctions (instead of the more localized Wannier functions) to describe the electrons in atoms; localization is modeled again through a second adjustable parameter entering a localization exponential term. Besides, Althaus et al. do not model the true monoclinic cell of As$_2$Se$_3$, but rather an orthorhombic cell; this fact is supported by the small deviation of $\beta$ from 90º. I admite that these approximations sound reasonable, and could probably be the only options at the time their paper was published, but they could also be the origin of the discrepancies with this work. On the other hand, Tarnow et al. \cite{Tarnow-86a} used DFT, but with a different flavor for the exchange-correlation functional, namely LDA, in their calculations. Since LDA and GGA should give comparable results for the electronic structure, I think that the origin of the discrepancies between this work and that by Tarnow et al. lies in the particular setup that these authors used. First, they performed convergence tests only to within 0.1~eV. This choice resulted in that they summed over the ZB only at $\vec k=\left(\frac 14, \frac 14, \frac 14\right)$ and their equivalent by symmetry, rather than over the 72 points used herein. More importantly, Tarnow \text{et al.} explicitly admit that interlayer interactions play an important role, although they do not include these corrections in their calculations. Finally, these authors did not model the actual monoclinic lattice, but rather an orthorhombic one. 

The uppermost valence band deserves a more detailed discussion. Althaus et al. argued that the group of 36 bands with energies just below the Fermi energy comprises the 48 Se p electrons and the 24 As p ones. The present results indicate that there is actually some hybridization, as for the previous groups of bands. At energies below about $-$1.2~eV the bands arise mainly from the hybridization of the p states of Se and As, but at higher energies the bands have mainly Se-s and p character. This feature is interesting, since it was early argued that a gap could exist within the topmost valence band between the lone-pair Se states and the As-Se bonding ones \cite{Zallen-71, Blabla-76}, a picture which likely arose from the relatively simple theoretical models employed by the authors and which was rejected by Althaus \textit{et al}. The results in Fig.~\ref{Fig_2}, on the other hand, point towards that the valence band forms by the juxtaposition of a group with predominant hybridization of Se and As p states (below -1.2~eV) and a second group, above that energy, with predominant hybridization of the Se s and p states. In other words, this work confirms the interpretation found by Althaus et al. in their preliminary study and suggest that the idea of bonding "lone pair" bands in As$_2$Se$_3$ is not correct. 

\begin{figure}[!h]
	\centering
	\includegraphics[width=0.75\columnwidth]{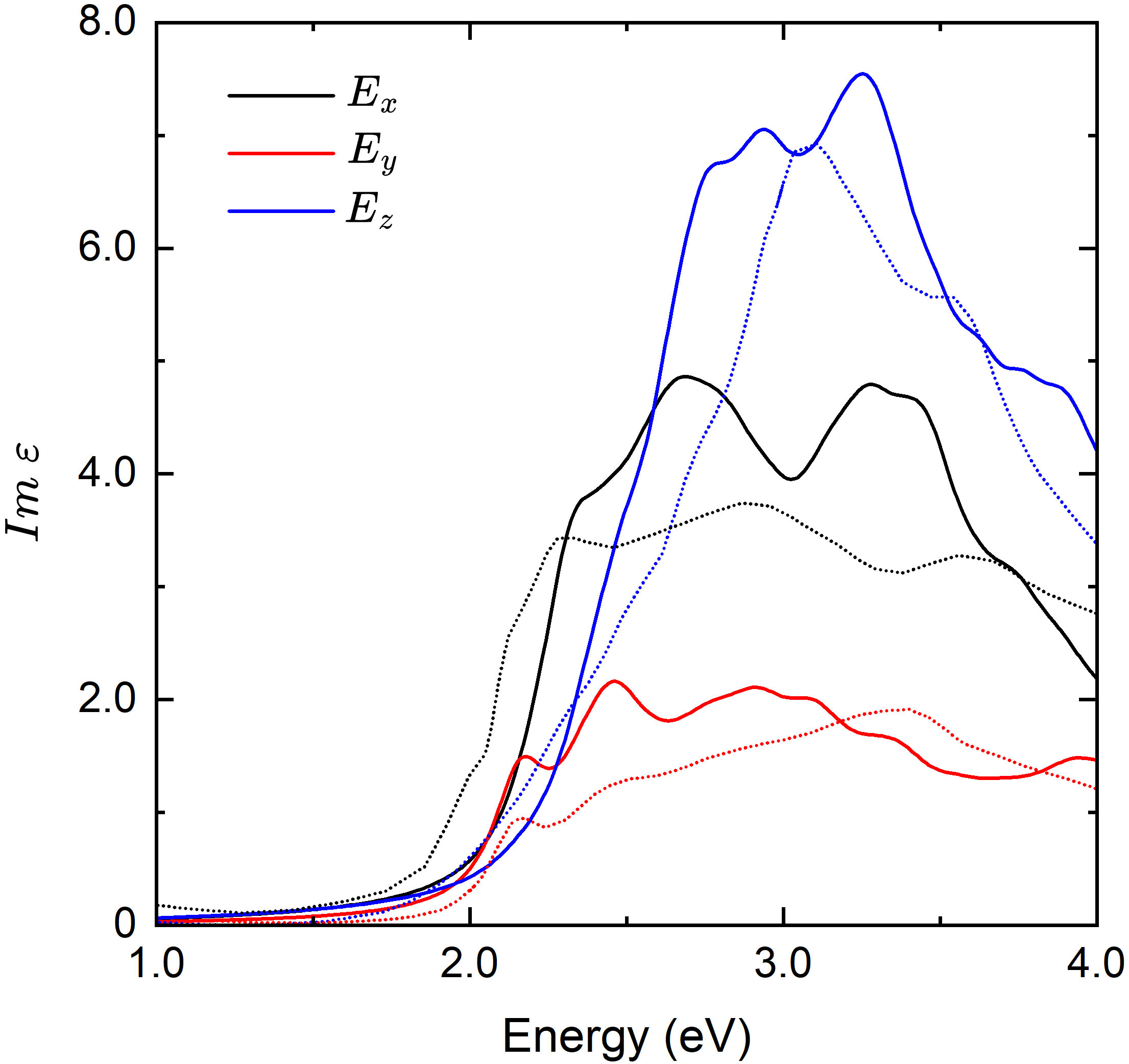}
	\caption{Computed imaginary part of the dielectric function $\kappa$ vs. energy for the three orientations of the electric field (continuous lines). Experimental data found in the literature are included as dotted lines.}
	\label{fig:Fig_5}
\end{figure}

As for the optical properties, there is only a very limited set of data available for comparison with absorption data computed within the G$_0$W$_0$+BSE formalism used here, to my knowledge. To check the validity of the calculations reported herein, Fig.~\ref{fig:Fig_5} plots the computed imaginary part of the dielectric function as a function of the energy for the three orientations of the electric field considered, together with experimental data reported by Tarnow et al. from data by Althaus et al. \cite{Tarnow-86b, Althaus-78}. The comparison of computed data with the experimental evidence available is quite good in what respects to the overall shape of the spectra (mainly in their respective onsets and maxima), the greater deviations occurring for $E \parallel E_z$, probably due to the need to include more bands in the resolution of the BSE. 

The values for the binding energies of the excitons from Table~\ref{table:excitons}, on the other hand, point out that these are actually Frenkel (i.e., localized, likely within the unit cell) excitons. An indirect estimation of the size of these excitons may be given from the projections in Fig.~\ref{fig:projections}. Fig.~S6 in the Supplement show details of these projections around the bandgap region. This figure evidence first that the projection of the topmost VB onto As atomic states takes on negligible values both at $\Gamma$ and at $B$ (as actually within the entire Brillouin zone), where the first and most intense excitons locate, respectively. As a consequence, the hole states for the excitons arising from this band are expected to locate near As atoms. On the other hand, whilst the projection of the bottommost CB onto Se-s states is certainly small, it takes on appreciable values for both Se-p and As atomic states, which indicates that, with the hole located at As atoms, there are nearby positions within the cell where a bond electron may locate. Thus, the findings of this work are consistent with the existence of localized excitons, with sizes of the order of the lattice parameters. The exception to this scenario is the most intense electron for the $E_x$ polarization, which exhibits a small binding energy and, therefore, is delocalized. 

To conclude, a brief mention to the nature of the optical bandgap. As I have already mentioned, there is some discrepancy about whether the onset of the optical absorption spectrum is due to direct (as pointed out by Zallen \cite{Zallen-76}) or indirect \cite{Blabla-76, Althaus-78}. In the light of the calculations presented in this work, it is quite clear that the gap of As$_2$Se$_3$ is actually indirect, but with close nearly-degenerated direct states at $\Gamma$ which may have have been fortuitously identified in some experiments. 

\section{Conclusions}
In this work, an exhaustive study of the electronic and optical properties of crystalline As$_2$Se$_3$ has been presented through an advanced \textit{ab initio} approach that combines DFT, G$_0$W$_0$ quasiparticle corrections, and the explicit treatment of electronic excitations via the Bethe--Salpeter equation (BSE). The results obtained resolve several long-standing controversies in the literature and provide a solid and unified theoretical framework for understanding this material.

First, the G$_0$W$_0$ calculations confirm that As$_2$Se$_3$ is an indirect-gap semiconductor, with a value of 2.31~eV, accompanied by nearly degenerate direct transitions at $\Gamma$. This feature explains why some experiments have interpreted the absorption edge as direct. Likewise, the detailed analysis of the electronic structure shows that the region near the top of the valence band is dominated by hybridized Se-s/p and As-p states, contradicting older models based on well-defined lone-pair states. Our results indicate that, rather than isolated non-bonding states, electronic hybridization and the weak dispersion near the VBM play a central role in the optical behavior of the material.

Regarding the optical properties, the spectra obtained through BSE reveal strong anisotropy in absorption and pronounced excitonic effects that do not appear within the independent-particle treatment. Bright excitons are identified for the three orientations of the electric field, several of them with significant binding energies, highlighting a strong electron--hole interaction. These excitons explain the features observed experimentally near the absorption onset and underscore the importance of many-body effects in crystalline As$_2$Se$_3$.

Finally, the comparison between the theoretical results and the available experimental data shows very good agreement in the overall shape of the spectra and in the relative positions of the absorption maxima, validating the methodology employed. Altogether, this study provides a complete and coherent description of As$_2$Se$_3$, clarifying the origin of its optical anisotropy, the role of electronic hybridization, and the excitonic nature of the absorption edge. These results reinforce the potential of this material for applications in anisotropic photonics and optoelectronic devices operating in the mid-infrared range.

\section{Acknowledgements}
The author acknowledges financial support through Grant No.\ PID2024-156352NB-I00, funded by MCIU/AEI/10.13039/

\noindent 501100011033/FEDER, UE, and
from Grant No.\ GR24022 funded by the Junta de
Extremadura (Spain) and by European Regional Development Fund (ERDF, “A way of making Europe”). The numerical calculations were carried out in the computing facilities of the Instituto de Computaci\'on
Cient\'{\i}fica Avanzada de Extremadura (ICCAEx).


\balance


\bibliography{as2se3} 
\bibliographystyle{elsarticle-num} 

\pagebreak
\widetext
\begin{center}\textbf{{\Large Supplement}}
\end{center}

\setcounter{equation}{0}
\setcounter{figure}{0}
\setcounter{table}{0}
\setcounter{page}{1}
\makeatletter
\renewcommand{\theequation}{S\arabic{equation}}
\renewcommand{\thefigure}{S\arabic{figure}}
\renewcommand{\bibnumfmt}[1]{[S#1]}
\renewcommand{\citenumfont}[1]{S#1}

\section*{Convergence tests}
\subsection*{G$_0$W$_0$ calculations}
In this section I provide convergence tests for the several parameters indicated in the main text. First, Figure S1a shows the variation of the gap with respect to the number of $\vec G$ vectors considered in the exchange part of the self-energy operator. These preliminary tests wer performed within the Hartree-Fock approximation to speed up the calculations; note that the convergence values do not change upon moving to G$_0$V$_0$. A converged value of 1929 reciprocal lattice vectors (corresponding to a kinetic energy cutoff of 9 Ry) was chosen. Analogously, Figure S1b shows the variation of the Hartree-Fock gap with respect of the number of points used to sample the first Brillouin zone; the 6$\times$6$\times$3 Monkhorst-Pack mesh was chosen for the subsequent calculations.

\begin{figure}[!h]
	\centering
	\begin{subfigure}{0.49\columnwidth}
		\centering
		\includegraphics[width=1.1\textwidth]{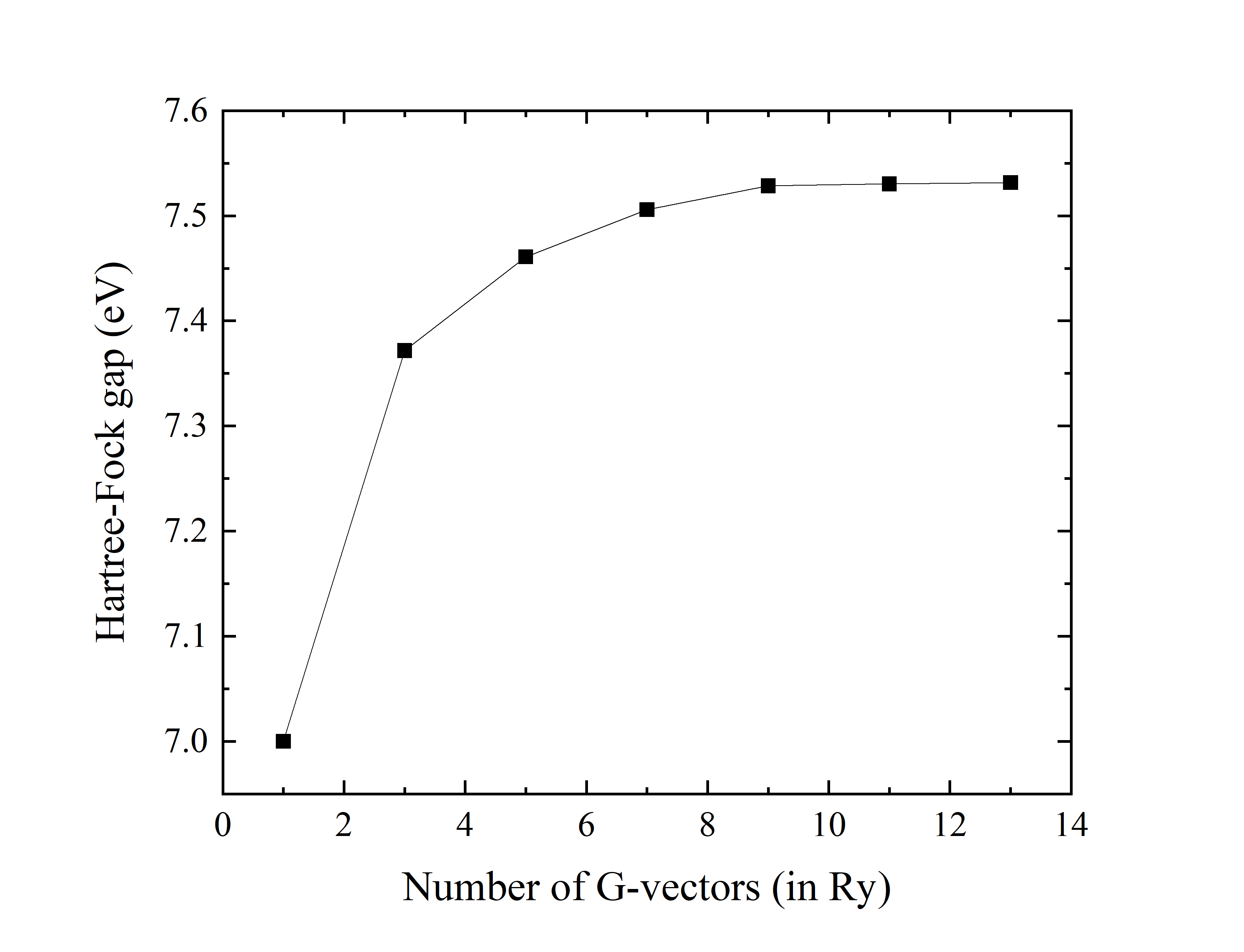}
		\caption{}
	\end{subfigure}
	\hfill
	\begin{subfigure}{0.49\columnwidth}
		\centering
		\includegraphics[width=1.1\textwidth]{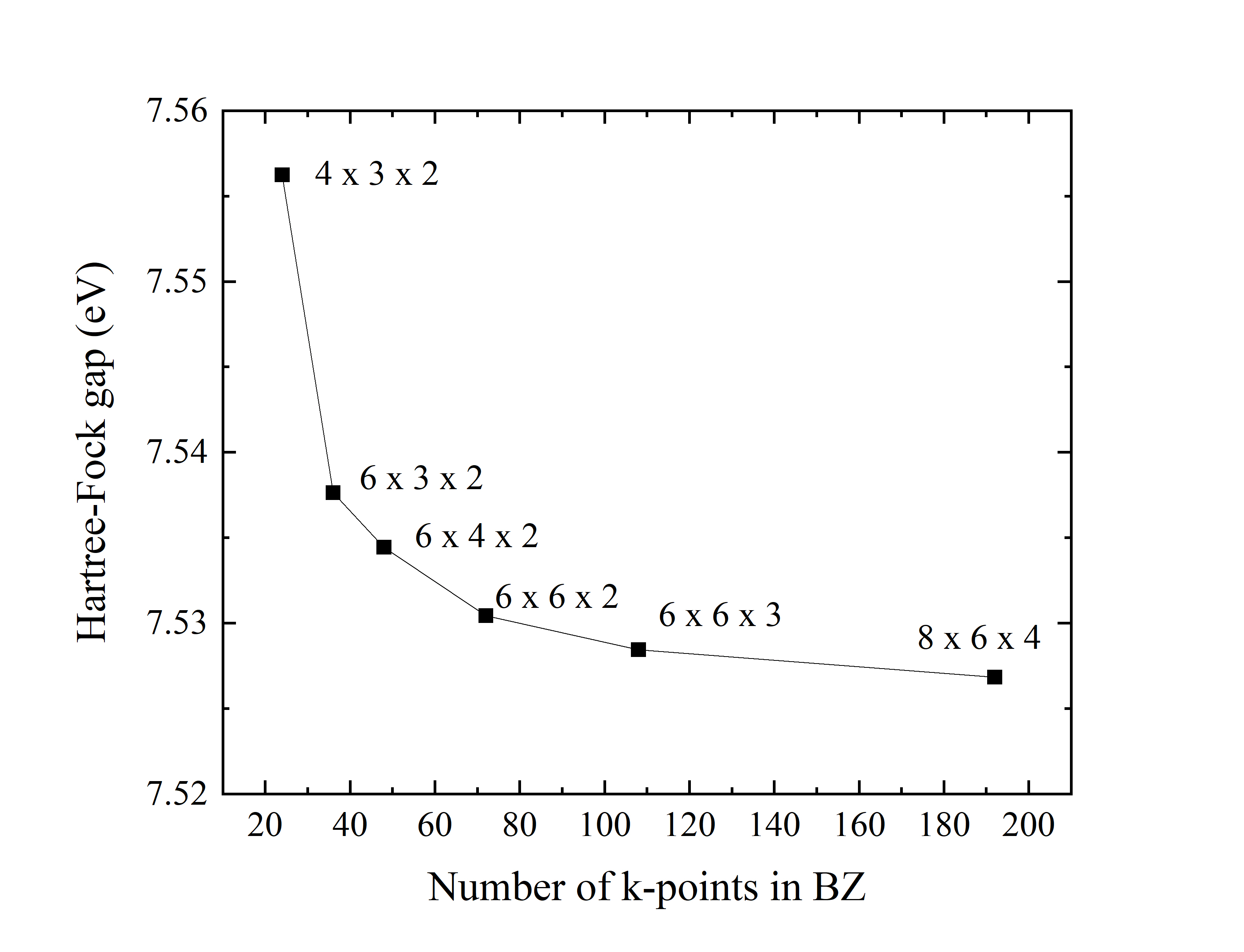}
		\caption{}
	\end{subfigure}
	\caption{\textit a) Study of the convergence of the Hartree-Fock gap with respect to the number of $\vec G$-vectors considered in the exchange part of the self-energy operator; \textit b) \textit{Idem} for the number of $k$ points sampling the first Brillouin zone; the number accompanying each symbol denotes the corresponding Monkhorst-Pack mesh.}
\end{figure}

Next, Figure S2a summarizes the convergence tests for the number of bands and of $\vec  G$ vectors entering in the calculation of the dynamical dielectric function, computed in this work within the PPA approximation. To speed up the calculations, these calculations were made by computing the correlation part of the self-energy from the valence bands only; for that reason, the converged values shown in Fig. S2a are not even close to the final converged bandgap. Note that these two variables are correlated in general, and their convergence cannot be studied separately. From this figure, I concluded that suitable values for the number of bands and $\vec G$-vectors yielding converged gap were 400 and 379 (equivalent to a kinetic energy cutoff of 3 Ry), respectively, as shown in the text. 

A final convergence study at the G$_0$W$_0$ level is required for the number of bands used to compute the correlation part of the self-energy operator, $\Sigma_c$. The results for this study is summarized in Figure S2b, from which one concludes that 300 bands yield converged values for $\Sigma_c$.

\begin{figure}[!h]
	\centering
	\begin{subfigure}{0.49\columnwidth}
		\centering
		\includegraphics[width=1.1\textwidth]{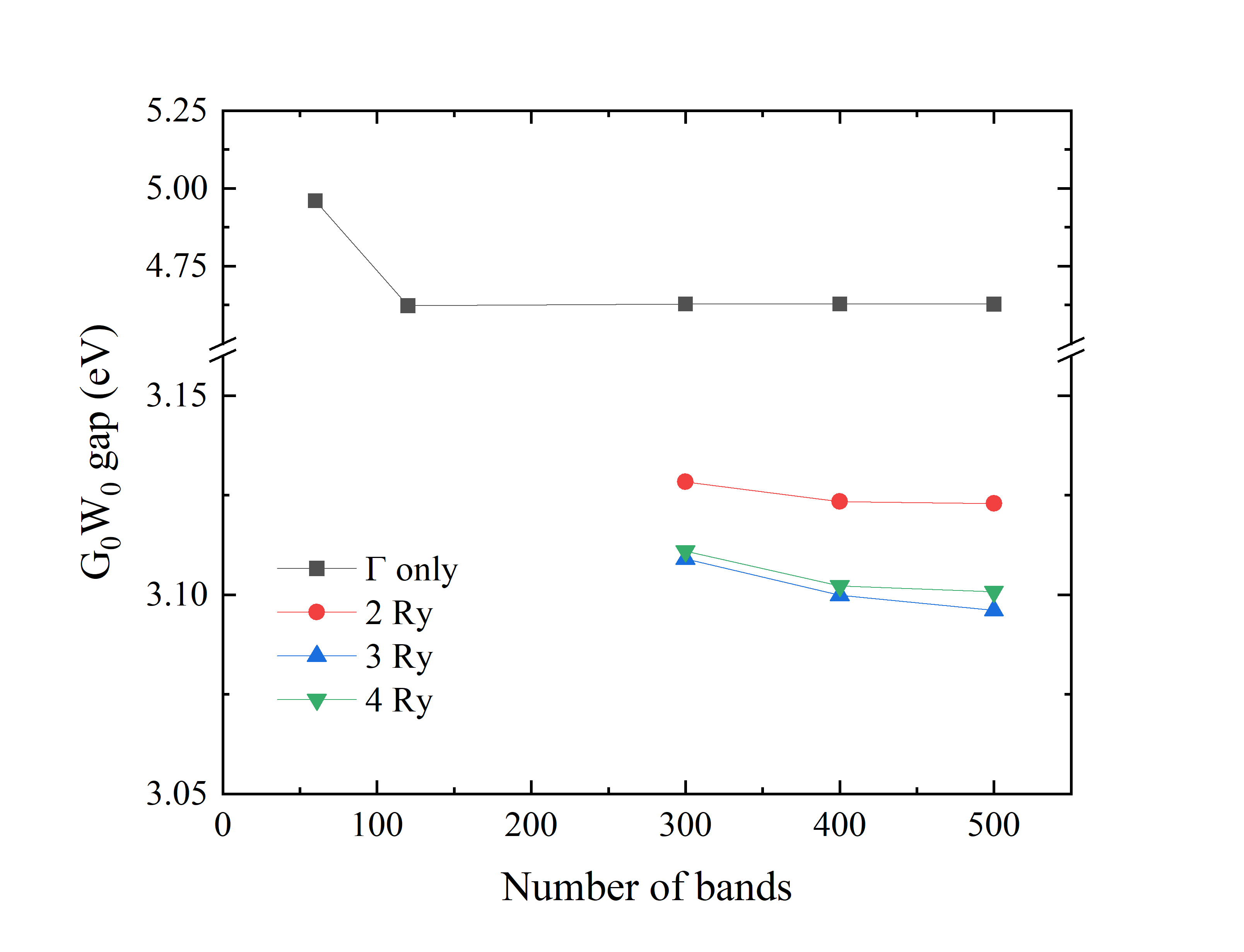}
		\caption{}
	\end{subfigure}
	\hfill
	\begin{subfigure}{0.49\columnwidth}
		\centering
		\includegraphics[width=1.1\textwidth]{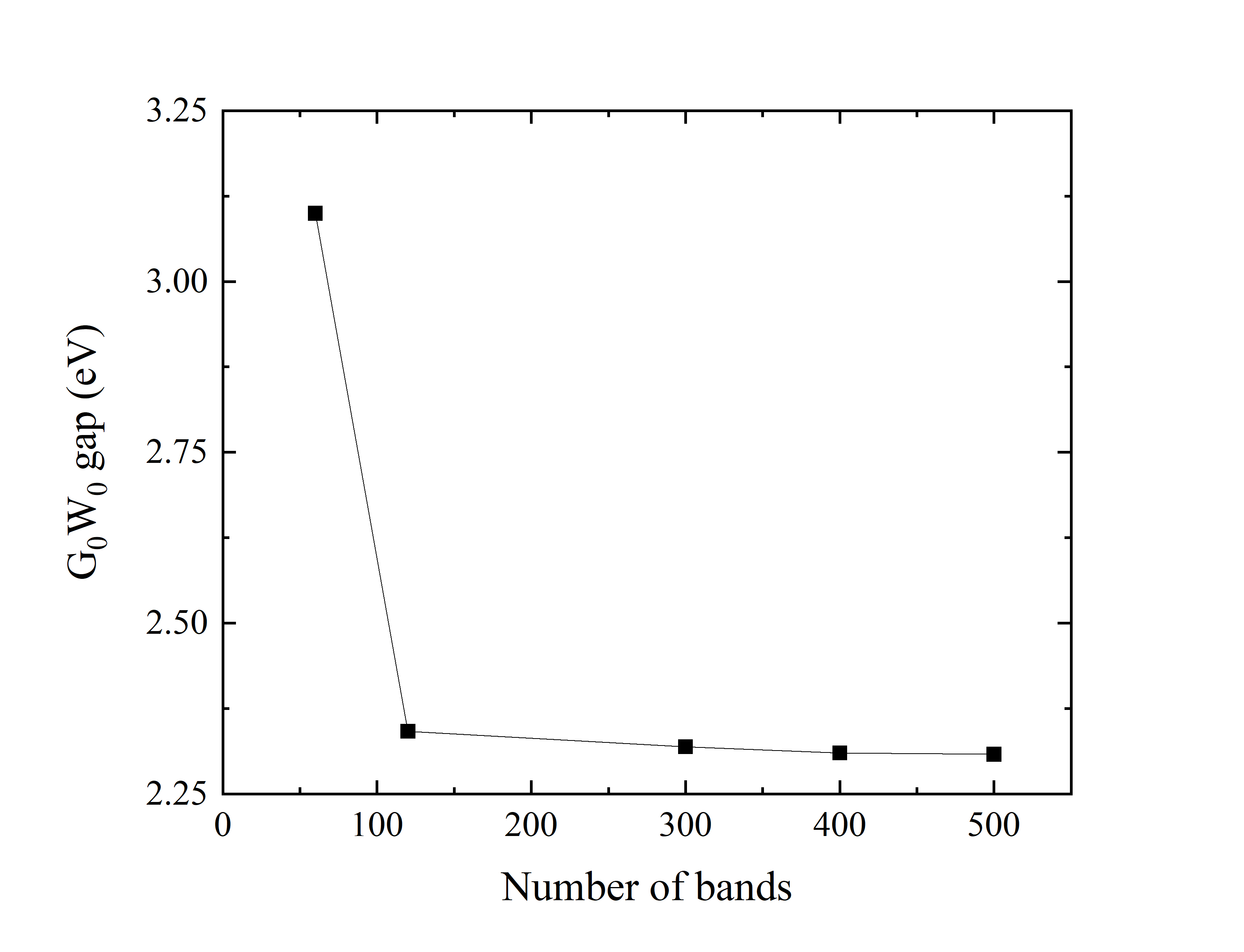}
		\caption{}
	\end{subfigure}
	\caption{\textit a) Study of the convergence of the bandgap with respect to the parameters, namely number of $\vec G$-vectors and of bands, required to compute the dielectric function; \textit b) \textit{Idem} for the number of bands entering in the calculation of the correlation part of the self-energy.}
\end{figure}

\subsection*{BSE calculations}
Next, one must carry out convergence studies on a number of variables to get accurate optical spectra, namely the number of $\vec G$-vectors required for the screened Coulomb potentials and the number of valence and conduction bands required to get a converged optical spectrum compatible with experiments. In this case, since I have studied the optical behavior for three different light polarizations, I have performed convergence tests for each spatial direction. Note that, in principle, one should study also the convergence of the Hartree potential with the number of $\vec G$-vectors considered in its calculation; however, this is a relatively light calculation, so that all the reciprocal lattice vectors from the DFT run were considered.

Regarding the number of reciprocal lattice vectors required to compute the screened potential, Figures S3 show the absorbance as function of the energy for the three orientations of the electric field and several numbers of $\vec G$-vectors; to speed up the calculations, the absorbance has been computed taking into account transitions between only 5 valence and 5 conduction bands. From these plots, one observes that converged spectra are obtained already for 83 $\vec G$-vectors (corresponding approximately to a cutoff of 1 Ry, as indicated in the text), regardless of the electric field orientation. Aditionally, Table S1 shows the stability of the first (i. e., lowest-energy) exciton with respect to these number of $\vec G$-vectors in terms of its energy and oscillation strength (related to the exciton character, i. e., bright or dark), again for the three orientations of the electric field considered. I have been reluctant to record the binding energy in Table S1, as it does not deal with the fully converged spectrum. As this table demonstrates, 1 Ry is a suitable choice for the number of $\vec G$-vectors required for the aforementioned calculations. 

\begin{figure}[!h]
	\centering
	\begin{subfigure}{0.49\columnwidth}
		\centering
		\includegraphics[width=1.1\textwidth]{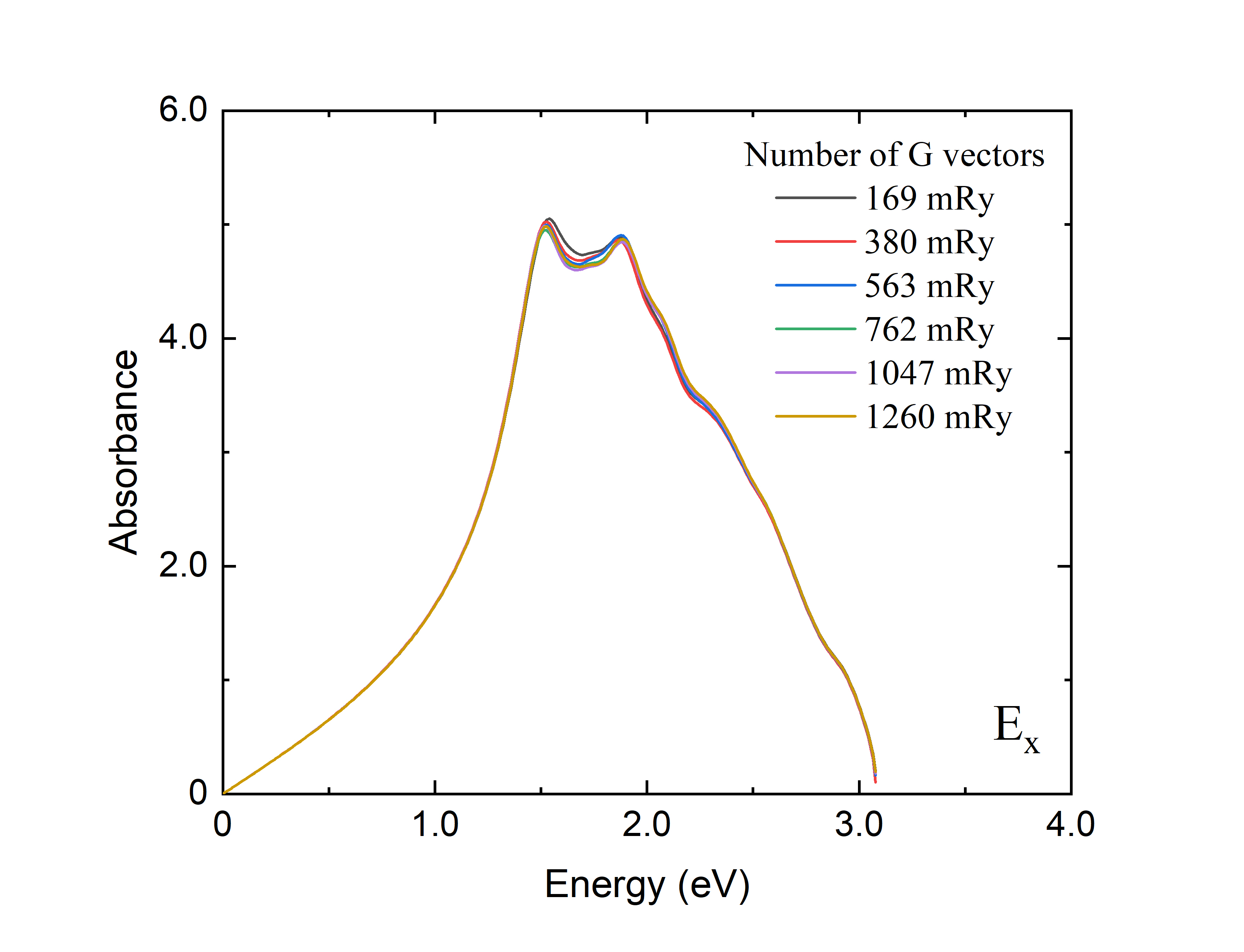}
		\caption{$E_x$}
	\end{subfigure}
	\hfill     
	\begin{subfigure}{0.49\columnwidth}
		\centering
		\includegraphics[width=1.1\textwidth]{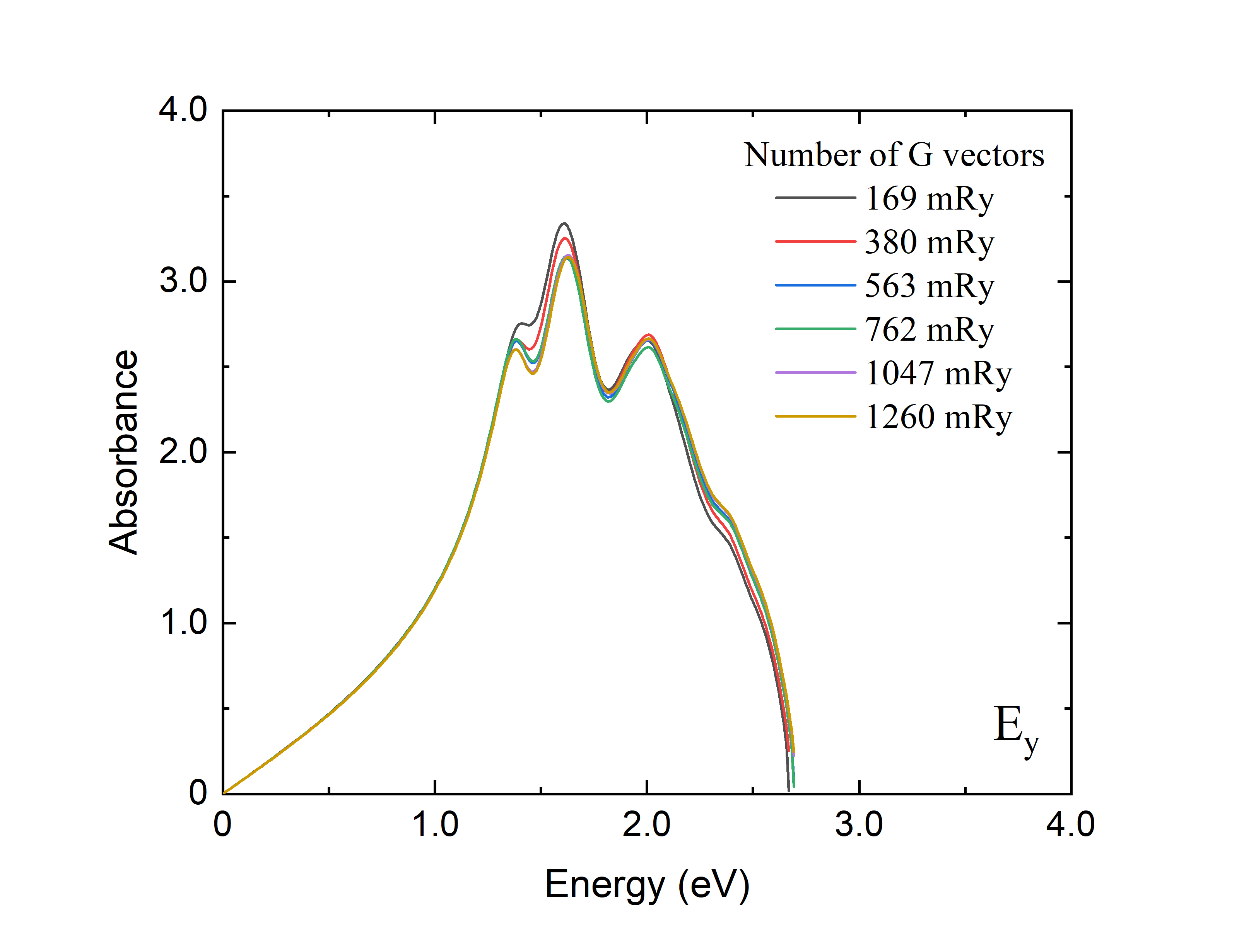}
		\caption{$E_y$}
	\end{subfigure}
	\qquad
	\begin{subfigure}{\columnwidth}
		\centering
		\includegraphics[width=0.55\textwidth]{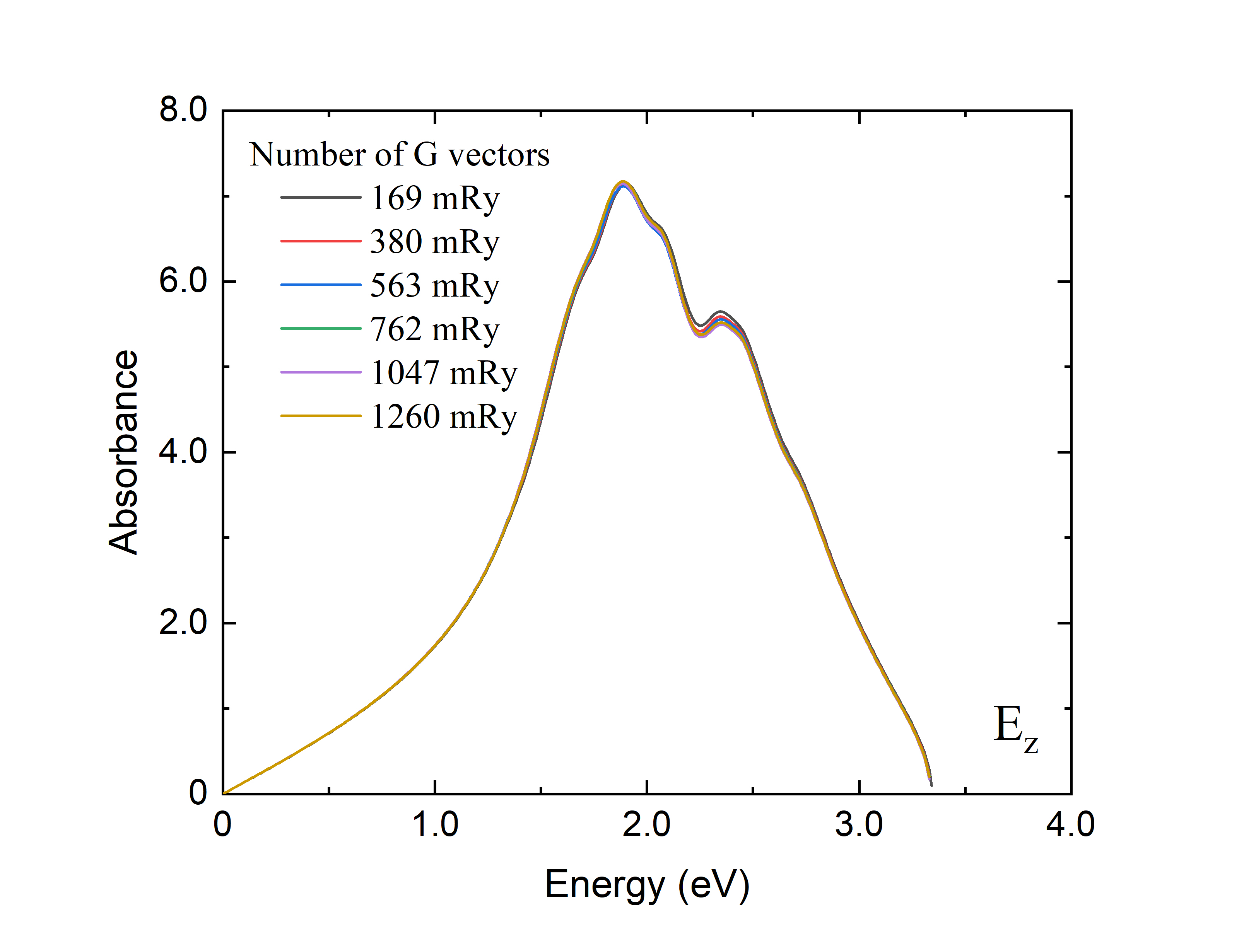}
		\caption{$E_z$}
	\end{subfigure}
	\caption{Partial absorption spectra (i. e., computed for only 10 bands) for several numbers of the $\vec G$-vectors entering the calculation of the Hartree and screened potentials for each orientation of the electric field.}
\end{figure}

\begin{table}[!h]
	\centering
	\renewcommand{\arraystretch}{1.5}
	\begin{tabular}{c | cc | cc | cc }
		\toprule
		\hline
		Number of $\vec G$	& \multicolumn{2}{c}{$\mathbf{E_x}$} &  \multicolumn{2}{c}{$\mathbf{E_y}$} & \multicolumn{2}{c}{$\mathbf{E_z}$} \\   \cline{2-7}			
		vectors (mRy) & $\varepsilon$ (eV) & Osc. strength & $\varepsilon$ (eV) & Osc. strength &  $\varepsilon$ (eV) & Osc. strength \\ \hline 
		169 & 1.463 & 0.475 & 1.402 & 1.0 & 1.401 & 0.281 \\ \hline
		380 & 1.465 & 0.541 & 1.396 & 1.0 & 1.405 & 0.350 \\ \hline
		563 & 1.464 & 0.389  & 1.397 & 1.0 & 1.403 & 0.347 \\ \hline
		762 & 1.465 & 0.476  & 1.396 & 1.0 & 1.407 & 0.205 \\ \hline
		1047 & 1.467 & 0.374  & 1.395 & 1.0 & 1.404 & 0.166 \\ \hline
		1260 & 1.468 & 0.422 & 1.396 & 1.0 & 1.405 & 0.212 \\ \hline
		\bottomrule
	\end{tabular}
	\caption{Energetic position and oscillation strength of the first bright exciton as function of the number of reciprocal lattice vectors used in the computation of the Hartree and screened Coulomb potential for the three directions of the electric fields considered in this work.}
\end{table}

Finally, one must converge the optical spectrum with respect to the number of valence and conduction bands yielding optical transitions; the results of this study is summarized in Figures S4, which plot the absorption spectra computed considering several numbers of valence (v) and conduction (c) bands. From these plots, one observes that the spectrum onset is quite reasonably converged for 6 valence bands, with the upper part of the spectrum changing obviously with the number of conduction bands. It is also apparent that 13 conduction bands suffice to get a reasonably converged absorption spectrum up to roughly 4 eV, which is the range plotted in Figs. 4 of the text. As for the excitonic properties, Table S2 shows the binding energy and oscillator strength of the first exciton for the three orientations of the electric field, shown that the choice of bands made in this work yields stable results.

\begin{figure}[!h]
	\centering
	\begin{subfigure}{0.49\columnwidth}
		\centering
		\includegraphics[width=1.1\textwidth]{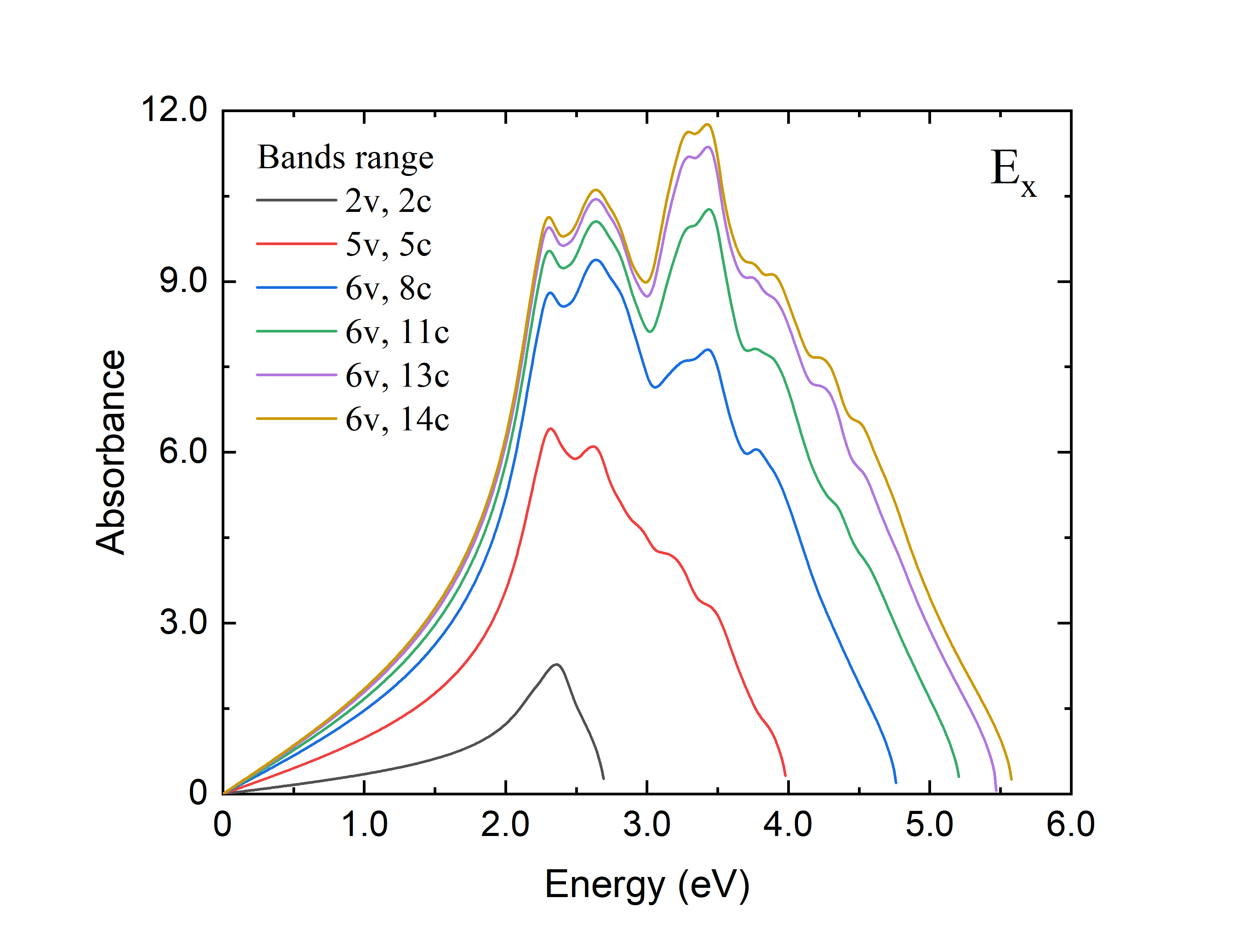}
		\caption{$E_x$}
	\end{subfigure}
	\hfill     
	\begin{subfigure}{0.49\columnwidth}
		\centering
		\includegraphics[width=1.1\textwidth]{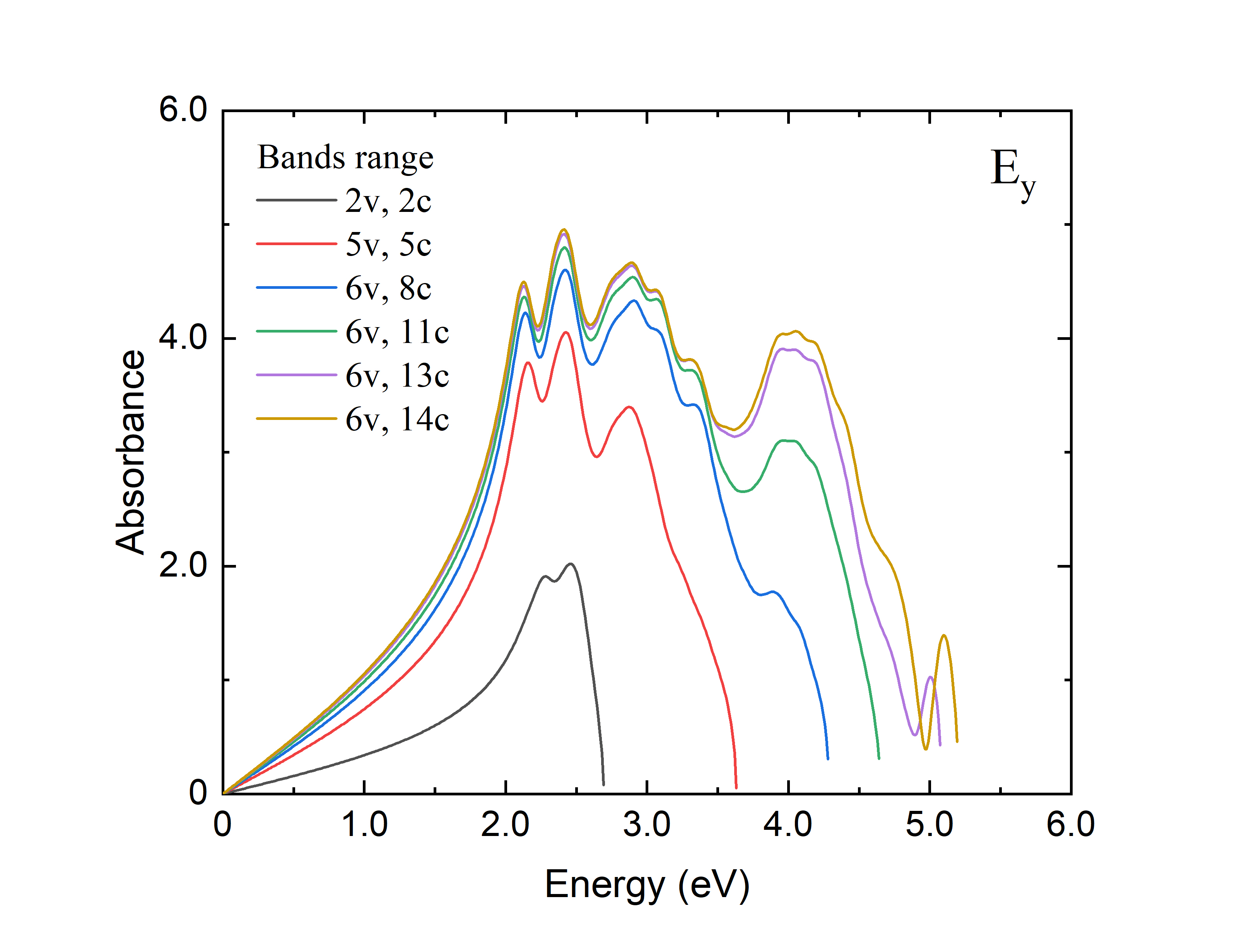}
		\caption{$E_y$}
	\end{subfigure}
	\qquad
	\begin{subfigure}{\columnwidth}
		\centering
		\includegraphics[width=0.55\textwidth]{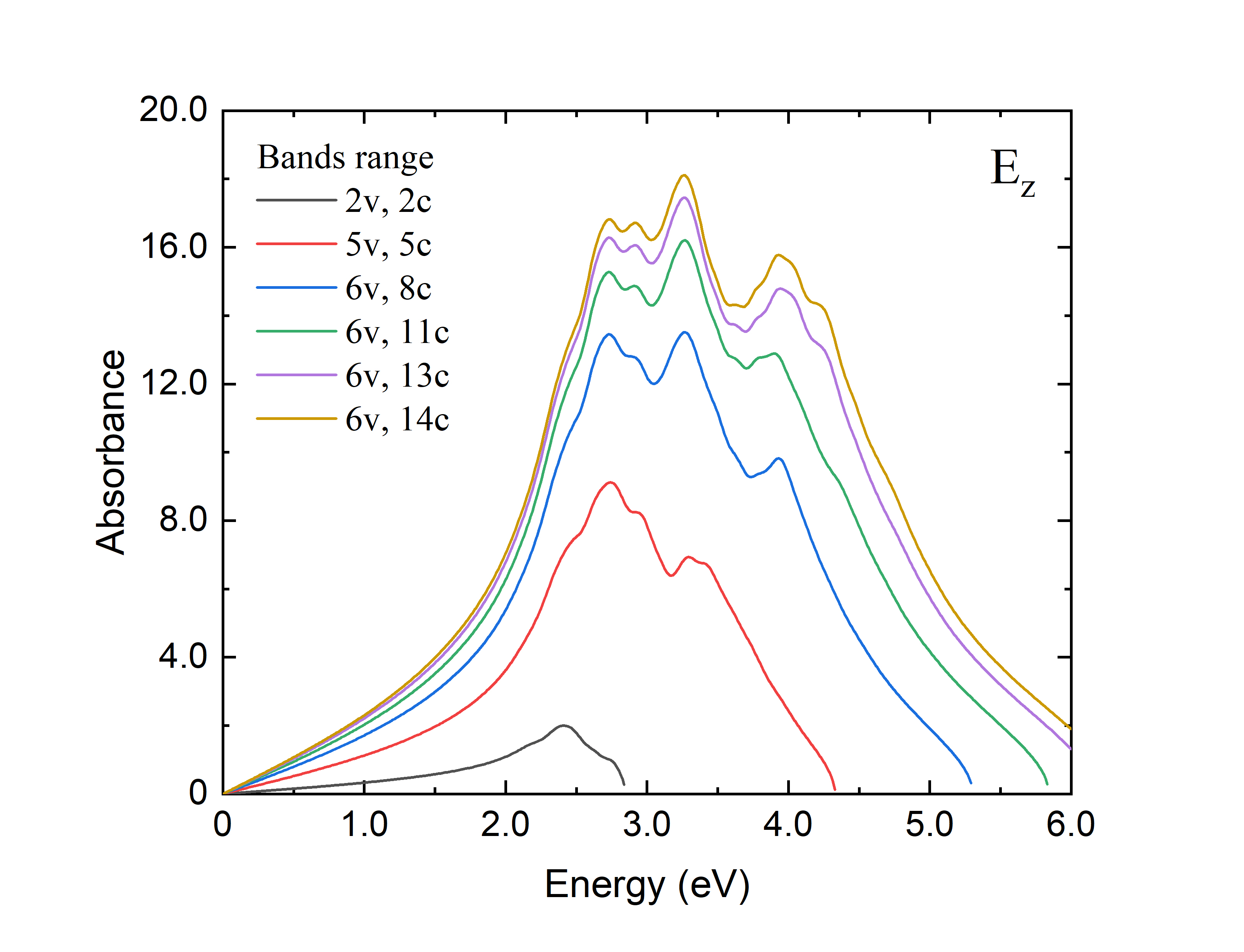}
		\caption{$E_z$}
	\end{subfigure}
	\caption{Partial absorption spectra (i. e., computed for only 10 bands) for several numbers of the $\vec G$-vectors entering the calculation of the Hartree and screened potentials for each orientation of the electric field.}
\end{figure}

\begin{table}[!h]
	\centering
	\renewcommand{\arraystretch}{1.5}
	\begin{tabular}{c | cc | cc | cc }
		\toprule
		\hline
		Number of $\vec G$	& \multicolumn{2}{c}{$\mathbf{E_x}$} &  \multicolumn{2}{c}{$\mathbf{E_y}$} & \multicolumn{2}{c}{$\mathbf{E_z}$} \\   \cline{2-7}			
		vectors (mRy) & $\varepsilon$ (eV) & Osc. strength & $\varepsilon$ (eV) & Osc. strength &  $\varepsilon$ (eV) & Osc. strength \\ \hline 
		2v, 2c &  0.1074 &  0.1882 & 0.10733 &  0.3061 & 0.1073 & 0.3061 \\ \hline
		5v, 5c &  0.0904 &  0.3151 & 0.1458 & 0.7147 & 0.1402 & 0.1971 \\ \hline
		6v, 8c &  0.0957 &  0.3603  & 0.15337 & 1.0 & 0.1509 &  0.2028 \\ \hline
		6v, 11c &  0.0980 &  0.4108 & 0.15783 & 1.0 & 0.1558 &  0.1585 \\ \hline
		6v, 13c &  0.0989 &  0.3989 & 0.15949 & 1.0 & 0.1574 &  0.1626 \\ \hline
		6v, 14c &  0.0992 &  0.3896 & 0.16 & 1.0 & 0.1583 &  0.1624 \\ \hline
		\bottomrule
	\end{tabular}
	\caption{Binding energy and oscillation strength of the first bright exciton as function of the number of bands used to compute the BSE absorption spectrum for the three directions of the electric fields considered in this work.}
\end{table}

\begin{figure}[!h]
	\centering
	\begin{subfigure}{0.49\columnwidth}
		\centering
		\includegraphics[width=\textwidth]{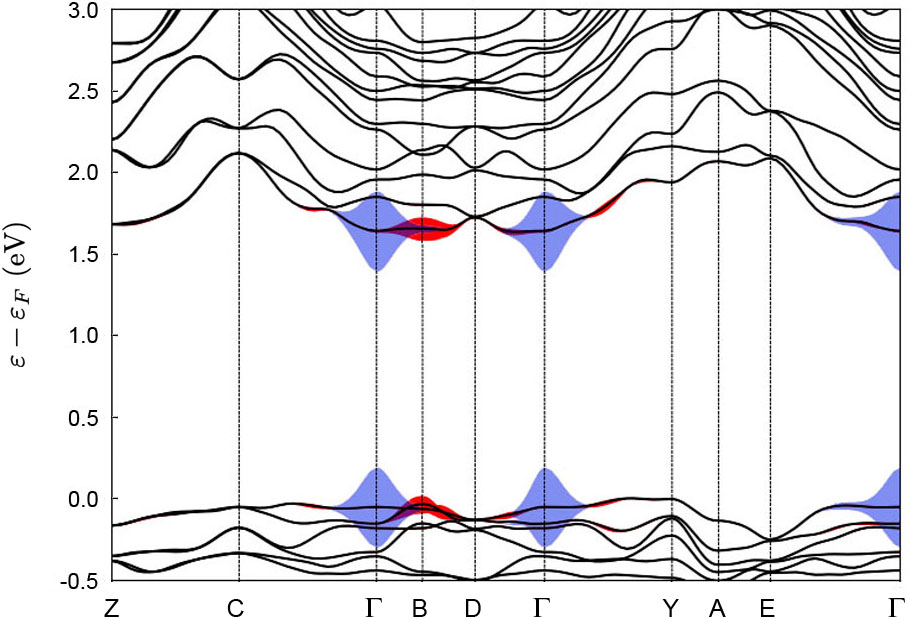}
		\caption{$E_x$}
	\end{subfigure}
	\hfill     
	\begin{subfigure}{0.49\columnwidth}
		\centering
		\includegraphics[width=\textwidth]{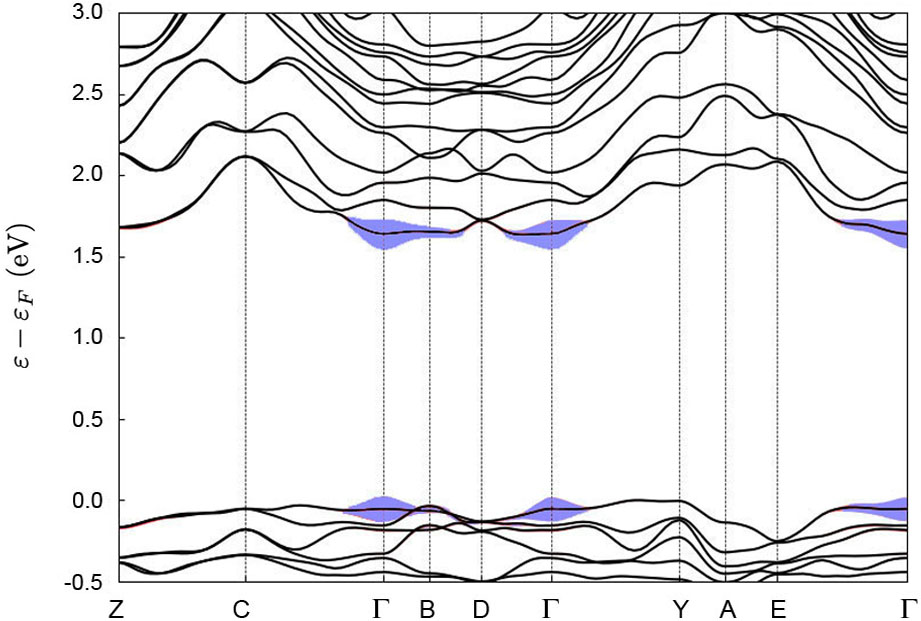}
		\caption{$E_y$}
		\label{fig:Exciton_Ey}
	\end{subfigure}
	\qquad
	\begin{subfigure}{\columnwidth}
		\centering
		\includegraphics[width=0.48\textwidth]{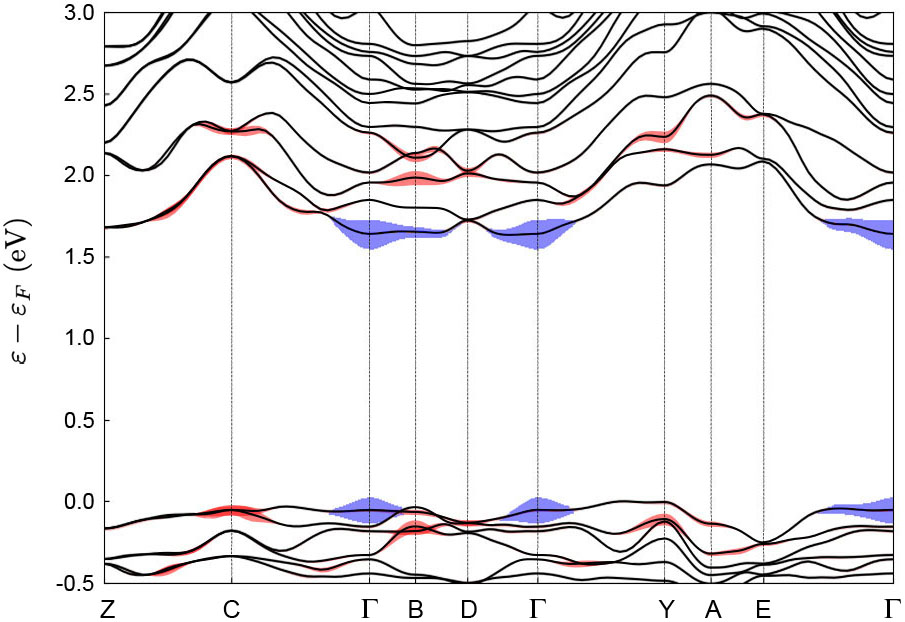}
		\caption{$E_z$}
	\end{subfigure}
	\caption{Plot of the regions contributing to the first (blue) and most intense (in red) excitons for the three orientations of the electric field. Note that, for $E \parallel E_y$, the first exciton is also the most intense.}
\end{figure}

\begin{figure}[!h]
	\centering
	\begin{subfigure}{0.49\columnwidth}
		\centering
		\includegraphics[width=\textwidth]{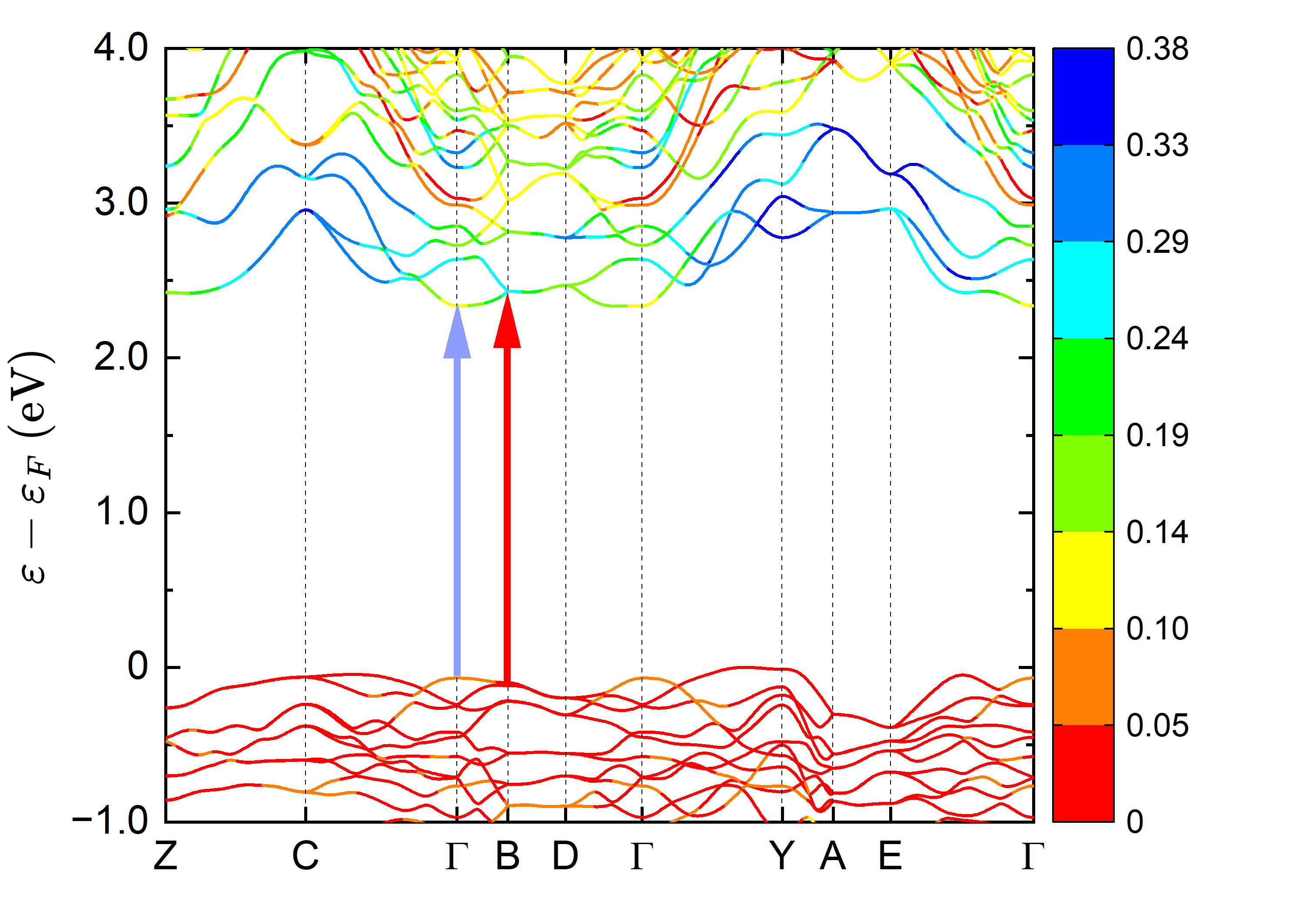}
		\caption{As-s}
	\end{subfigure}
	\hfill     
	\begin{subfigure}{0.49\columnwidth}
		\centering
		\includegraphics[width=\textwidth]{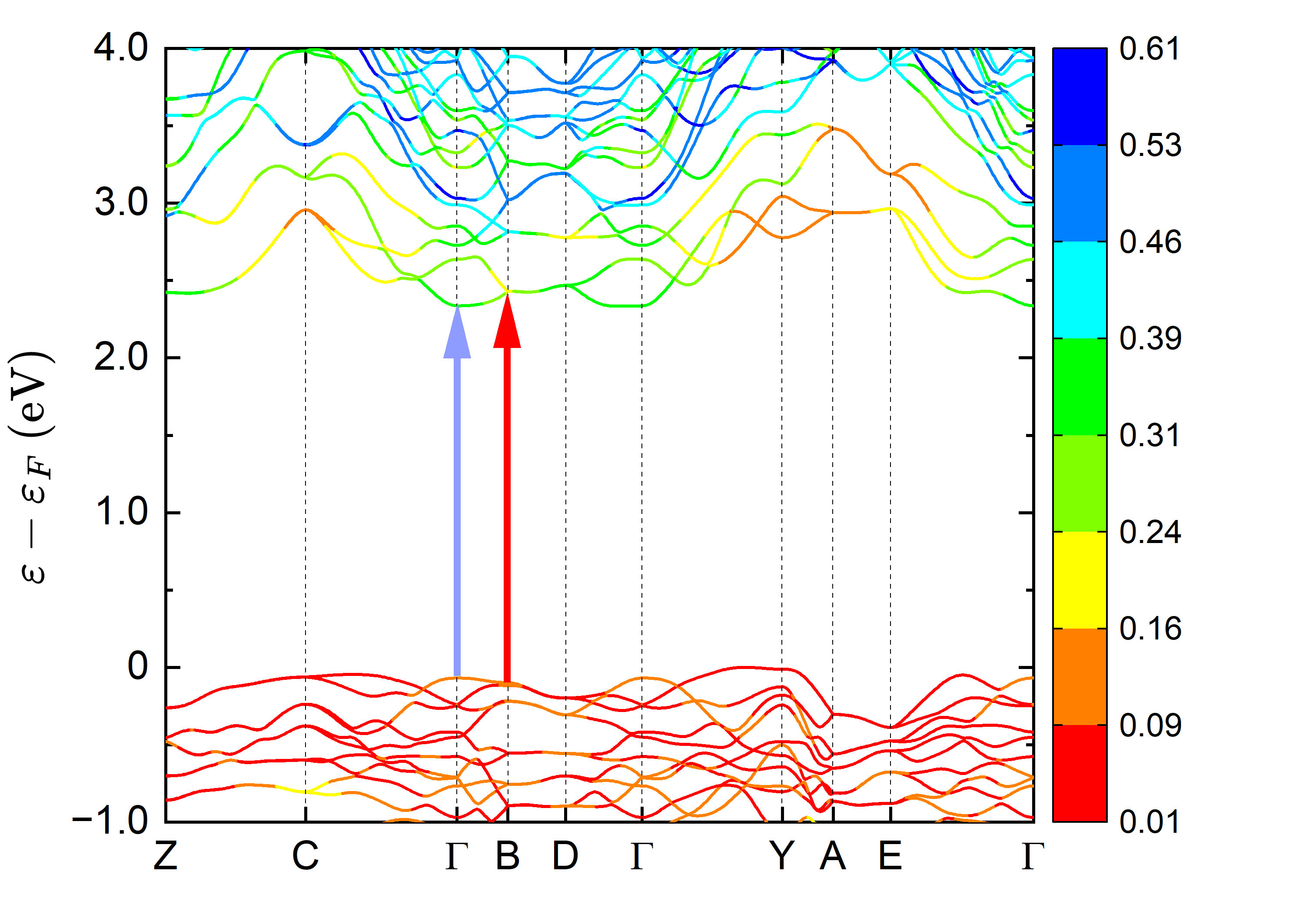}
		\caption{As-p}
	\end{subfigure}
	\qquad
	\begin{subfigure}{0.49\columnwidth}
		\centering
		\includegraphics[width=\textwidth]{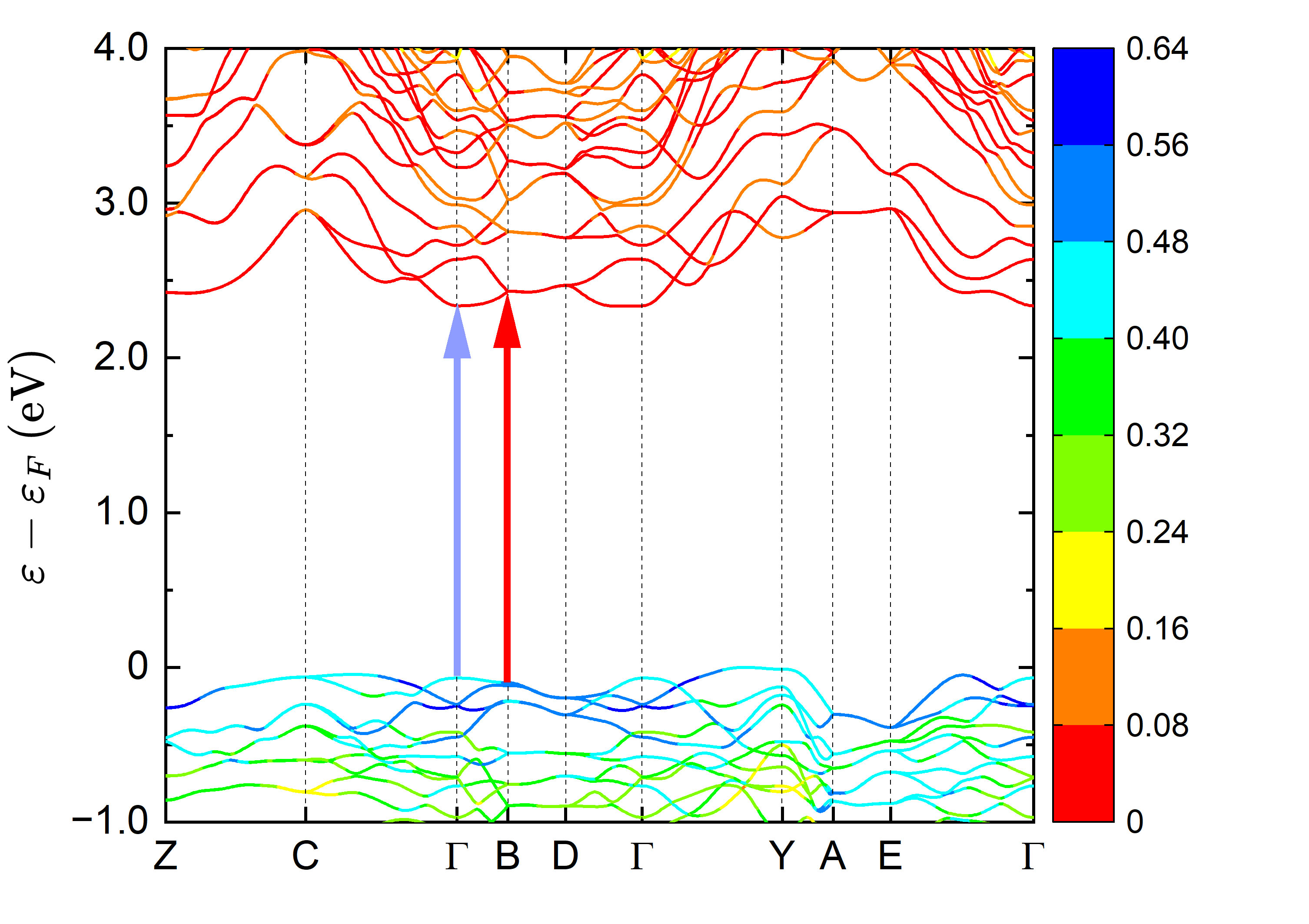}
		\caption{Se-s}
	\end{subfigure}
	\hfill     
	\begin{subfigure}{0.49\columnwidth}
		\centering
		\includegraphics[width=\textwidth]{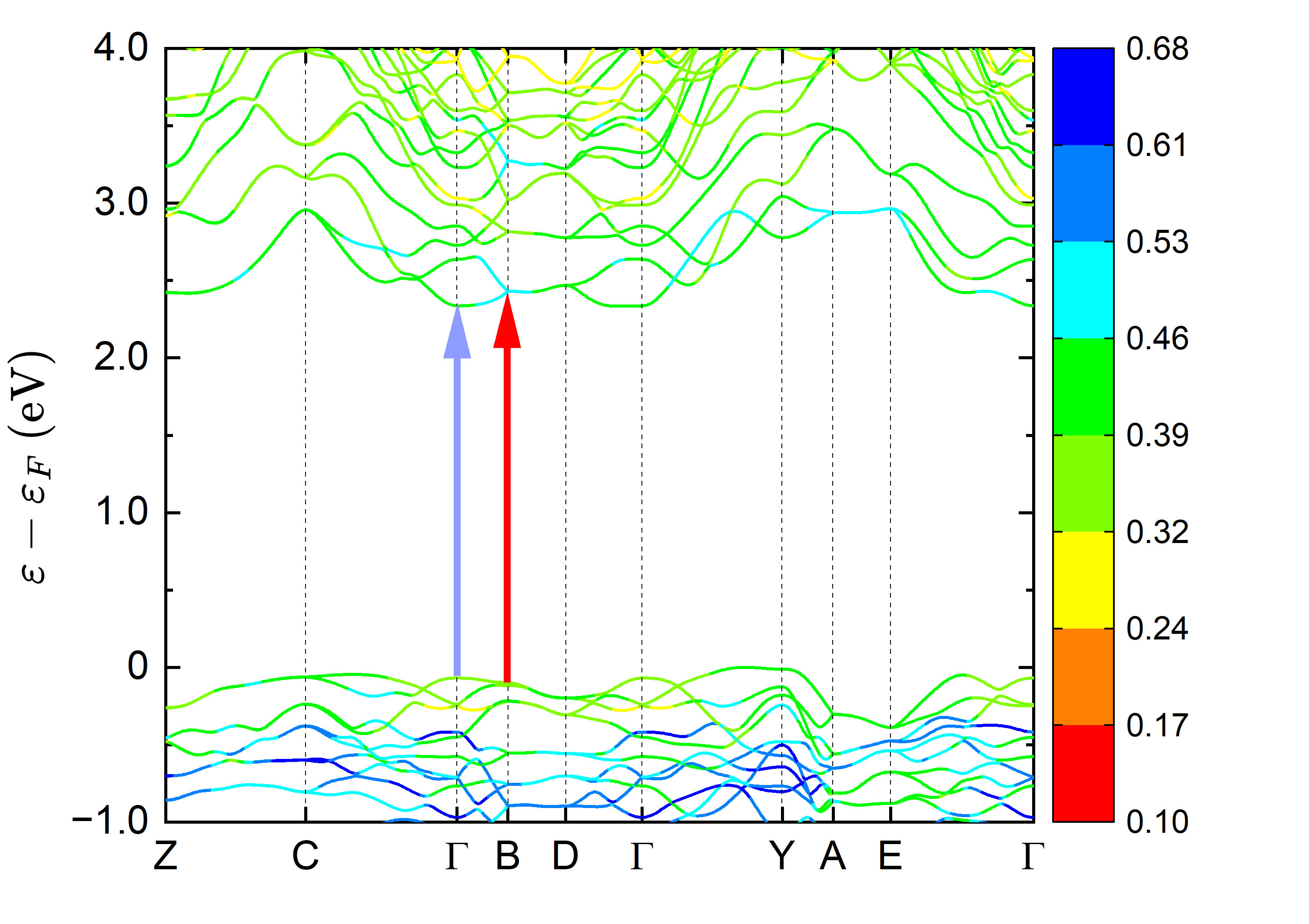}
		\caption{Se-p}
	\end{subfigure}
	\caption{Details of the projection of the band structure onto atomic states of As and Se (See Fig. 5 in the text) nearby the band gap. The vertical arrows schematize the positions within the BZ of the first (blue) and most intense (red) excitons.}
\end{figure}

\end{document}